\documentclass[11pt]{article}
\usepackage[a4paper,margin=1in]{geometry}

\usepackage{amsmath}
\usepackage{mathtools}

\usepackage[T1]{fontenc}
\usepackage[utf8]{inputenc}
\usepackage{mathptmx}
\usepackage{newtxtext}

\usepackage{array}
\usepackage{subcaption}
\usepackage{graphicx}
\usepackage{newunicodechar}
\newunicodechar{χ}{\ensuremath{\chi}}
\newunicodechar{≥}{\ensuremath{\geq}}
\newunicodechar{−}{\ensuremath{-}}

\date{} % no date

\newcommand{\keywords}[1]{\par\noindent\textbf{Keywords: }#1\par}

\usepackage{placeins}
\usepackage{booktabs}
\usepackage{multirow}
\usepackage{hyperref}
\usepackage[numbers]{natbib}
\usepackage{setspace}
\usepackage{authblk}
\usepackage{pdflscape}
\usepackage{soul}
\usepackage{xcolor}
\usepackage[left,mathlines]{lineno}

\providecommand{\keywords}[1]{\par\noindent\textbf{Keywords: }#1\par}

\title{Embryonic Exposure to VPA Influences Chick Vocalisations: A Computational Study}

\author[1]{Antonella M.\,C.\ Torrisi}
\author[1]{Inês Nolasco}
\author[2]{Paola Sgadò\thanks{Authors for correspondence: Paola Sgadò (\href{mailto:paola.sgado@unitn.it}{paola.sgado@unitn.it}), Elisabetta Versace (\href{mailto:e.versace@qmul.ac.uk}{e.versace@qmul.ac.uk}), Emmanouil Benetos (\href{mailto:e.benetos@qmul.ac.uk}{e.benetos@qmul.ac.uk})}}
\author[3]{Elisabetta Versace*}
\author[1]{Emmanouil Benetos*}

\affil[1]{Centre for Digital Music, Queen Mary University of London, London, UK}
\affil[2]{Centre for Mind/Brain Sciences, University of Trento, Rovereto, Italy}
\affil[3]{School of Biological and Behavioural Sciences, Centre for Brain and Behaviour, Queen Mary University of London, London, UK}

\begin{document}

\modulolinenumbers[1] 
\setlength\linenumbersep{9pt} 
\renewcommand\linenumberfont{\normalfont\scriptsize\sffamily}

\maketitle

\begin{abstract}

In young animals like poultry chicks (\textit{Gallus gallus}), vocalisations convey information about affective and behavioural states. Traditional approaches to vocalisation analysis, relying on manual annotation and predefined categories, introduce biases, limit scalability, and fail to capture the full complexity of vocal repertoires. We introduce a computational framework for automated detection, acoustic feature extraction, and unsupervised learning of chick vocalisations. Applying this framework to a dataset of newly hatched chicks, we identified two primary vocal clusters. We then tested our computational framework on an independent dataset of chicks exposed during embryonic development to vehicle or Valproic Acid (VPA), a compound that disrupts neural development and is linked to autistic-like symptoms. Clustering analysis on the experimental dataset confirmed two primary vocal clusters and revealed systematic differences between groups. VPA-exposed chicks showed an altered repertoire, with a relative increase in softer calls. VPA differentially affected call clusters, modulating temporal, frequency, and energy domain features. Overall, VPA-exposed chicks produced vocalisations with shorter duration, reduced pitch variability, and modified energy profiles, with the strongest alterations observed in louder calls. This study provides a computational framework for analysing chick vocalisations, advancing knowledge of early-life communication in typical and atypical vocal development.

\end{abstract}


\section{Introduction}
% general context
From the early stages of life, vocalisations play a crucial role in communication, facilitating social interactions \cite{edgar2016influences}, conveying affective states \cite{andrew1969intracranial} and mediating behavioural outcomes \cite{briefer2020coding}. In young domestic chicks (\textit{Gallus gallus}), vocalisations are an indicator of psychophysiological states \cite{manteuffel2004vocalization}, welfare conditions \cite{fontana2015innovative, herborn2020spectral} and developmental health \cite{laurijs2021vocalisations}. Their quantitative analysis is therefore important for both scientific investigation and practical applications. Chickens have a pivotal socioeconomic value in industrial farming \cite{bist2024sustainable,ozenturk2024robotics}. They are also a model in neuroscience for early socio-cognitive abilities \cite{di2017filial,mccabe2019visual,nicol2015behavioural}, in typical and atypical developmental processes \cite{sgado2018embryonic,nishigori2013impaired,versace2019transient}. % Problem specific / the gap we filled
Although chick vocalisations have been studied for decades \cite{collias1953spectrographic}, their classification has traditionally relied on manual annotation and predefined categories based on visual inspection of spectrograms \cite{marx2001vocalisation}. This qualitative approach is shaped by anthropocentric interpretations which may overlook biologically relevant aspects of the signal.
Recent attempts using supervised methods, such as convolutional neural networks, have made progress but fail to characterise calls that do not match the predefined types \cite{thomas2023using}. Efforts to describe chick calls through acoustic features \cite{collins2024sound, de2025automated} have similarly remained anchored to \textit{a priori} categories rather than empirically derived vocal repertoires. A data-driven and automated system for analysing chick vocalisations is therefore still lacking.
% our proposal/contribution
To address these limitations, we developed a computational framework that combines automated call detection, acoustic feature analysis, and clustering of individual calls to map the vocal repertoire of young chicks. Using an initial dataset of untreated newly hatched chicks, we identified vocal categories and patterns of acoustic variation based on 20 interpretable features extracted from the temporal, frequency, and energy domains. We then used the framework to compare the vocalisations of chicks exposed to Valproic Acid (VPA) or vehicle-injected (controls), from a subset of a dataset originally collected to study the effect of embryonic VPA exposure on early social behaviour \cite{lorenzi2019embryonic}. Unlike supervised approaches, our framework characterises chick vocalisations without predefined human labels and can process secondary data, reducing the need for dedicated acoustic data collection.

\subsection{Research contributions}

This study offers a primarily methodological contribution, demonstrated through a biological application. First, we developed a computational framework that addresses current methodological limitations in vocal analysis of chicks by reducing reliance on predefined labels and deriving the vocal repertoire directly from the data rather than assigning calls to predefined categories. The framework was designed to analyse secondary data, including audio extracted from behavioural videos not originally acquired for acoustic analysis.
Second, we applied an analytical framework to call-level data to study the impact of embryonic exposure to VPA on chick vocalisations, comparing VPA-exposed and control chicks through three main analyses: \textit{(i)} call production, \textit{(ii)} repertoire composition through clustering-based analysis, and \textit{(iii)} feature analysis. By combining feature and clustering analysis, this approach can identify vocal markers of neurodevelopmental disruption at both the repertoire and the feature level.

\section{Background and state of the art}

\subsection{Computational frameworks for animal vocalisation analysis}

Traditional bioacoustic research, including call detection, feature extraction, and classification, has been constrained by labour-intensive manual processes that are difficult to scale to large datasets \cite{stowell2016bird}. Computational approaches have begun addressing this limitation\cite{stowell2022computational, xie2023review}. Supervised methods, from convolutional networks to attention-based architectures and large models for multi-species recognition tasks, have improved the automatic detection and classification of animal vocalisations by learning context-aware representations and capturing complex acoustic dependencies \cite{noumida2022multi, robinson2024naturelm}. However, reliance on supervised learning pipelines based on human-labelled data still imposes an annotation burden \cite{Stowell2014AutomaticLC} and can propagate experimenter biases into the resulting models \cite{stowell2022computational}.
Alternative approaches, such as semi-supervised and few-shot learning methods \cite{van2020survey,nolasco2023learning}, are less dependent on large labelled datasets, but remain constrained by observer-dependent reference selection, which may limit the representation of acoustic variability. Recent hybrid approaches combine supervised recognition with clustering of residual audio \cite{grzywalski2025broiler}, but can be limited by interval-level detection and speech-pretrained embeddings that may not capture avian-specific vocal patterns. Self-supervised methods learn representations directly from unlabelled inputs \cite{moummad2024self}, but they still require labelled examples for downstream evaluation and thus impose data structure assumptions that may mask the intrinsic vocal repertoires \cite{goffinet2021low,sainburg2020finding}. 
Fully unsupervised approaches, by contrast, offer a more data-driven avenue to study natural acoustic categories by identifying acoustic categories directly from the statistical properties of the vocal data, without pre-imposed labels \cite{sainburg2021toward}. Further, operating without manual annotation, they scale well to large data volumes \cite{Stowell2014AutomaticLC}.
Fine-grained acoustic characterisation can complement these methods through the extraction of one-dimensional features via signal processing techniques \cite{elie2016vocal,wang2022joint}, providing a rich and interpretable representation of vocal categories \cite{fischer2017structural}.
While deep learning approaches can capture complex acoustic relationships \cite{stowell2022computational}, feature-based analyses remain preferable when understanding the underlying mechanisms is as important as classification accuracy \cite{goffinet2021low}. Such analyses can also reveal subtle acoustic distinctions that reflect internal states and serve as quantitative indicators of welfare and development \cite{briefer2020coding}.
 
\subsection{Application to a Valproic Acid model of neurodevelopmental disorders}

We tested our computational framework on a dataset collected to study the effect of embryonic exposure to Valproic Acid (VPA) on early social behaviour \cite{lorenzi2019embryonic}. Valproic Acid (VPA) is an anticonvulsant and mood stabiliser used to treat epilepsy, migraines, and bipolar disorder \cite{johannessen2003valproate}. Exposure to VPA during prenatal life interferes with the development of the social brain and increases the risk of developing ASD \cite{christensen2013prenatal}. In animal models, exposure to VPA during embryonic development recapitulates autistic-like symptoms, including social deficits, repetitive behaviours, and alterations in sensory processing \cite{kataoka2013autism, roullet2013utero, soares2026vocal}.
Several investigations have been conducted using rodents \cite{nicolini2018valproic} as well as chicks \cite{nishigori2013impaired, sgado2018embryonic, csillag2022avian} exposed to VPA during the embryonic phase. In chicks, exposure to VPA during development leads to impairments in social behaviour \cite{sgado2018embryonic, zachar2019valproate}, including a loss of early social predispositions \cite{sgado2018embryonic,lorenzi2019embryonic, adiletta2021spontaneous} and compromised aggregative behaviour \cite{nishigori2013impaired}. 
Interestingly, the neurobiological bases of some of these deficits are starting to be investigated \cite{adiletta2022embryonic}.
Research across species has revealed VPA-related vocal alterations. In marmosets, VPA exposure alters call type distributions \cite{uesaka2023automatic} and disrupts kinship-related vocal patterns during parent-infant interactions \cite{mimura2026altered}. However, a critical limitation of these studies is the high proportion of unclassified vocalisations, which may mask additional group differences. 
In rodents, prenatal VPA exposure reduces vocalisation rate and call complexity, with shorter calls, elevated peak frequency, and developmental shifts toward simpler call types, reflecting broad alterations in vocal repertoire \cite{moldrich2013inhibition, nicolini2018valproic, gzielo2020valproic}. In chicks, VPA exposure can reduce the number of vocalisations in threat-related contexts \cite{zachar2019valproate} and lower call frequency and amplitude \cite{nishigori2013impaired}. In zebra finches, VPA embryonic exposure disrupts song learning and vocal development, resulting in altered tutor imitation, delayed song crystallisation, and increased song entropy \cite{tewelde2025prenatal, soares2026vocal}. 
As a precocial species, chicks can be tested without the maternal care present in altricial models, such as rodents, and their vocalisations can be recorded under fully controlled conditions from the first hours of life \cite{sgado2018embryonic}. Moreover, the in-ovo embryonic development enables a more efficient control of pharmacological manipulation, eliminating any maternal interference. The fine-grained phonetic and acoustic structure of vocalisations in VPA-exposed chicks nonetheless remains understudied \cite{matsushima2024domestic}. A comprehensive, data-driven analysis of acoustic features is therefore essential for identifying vocal markers in translational neurodevelopmental research.

\section{Methods}

\begin{figure}[htbp]
 \centering
 \includegraphics[width=0.95\textwidth]{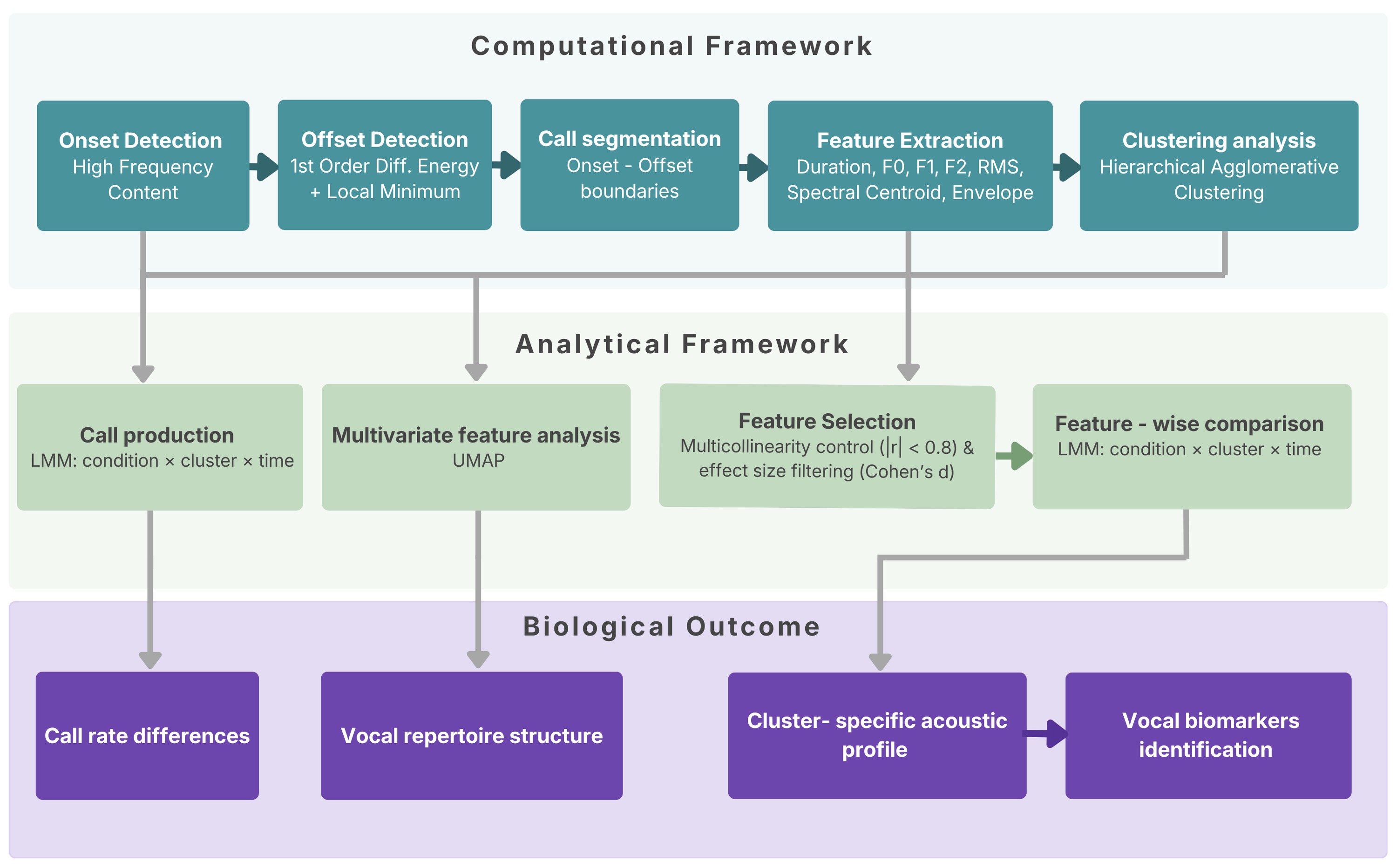}
\caption{
Computational (top), analytical (middle), and biological outcome (bottom) framework for call-level analysis of chick vocalisations. Horizontal arrows indicate processing steps, and vertical arrows information flow across layers.
}

 \label{fig:chick_vocalization_framework}
\end{figure}

The methods are organised around two core components (see Figure~\ref{fig:chick_vocalization_framework}): the computational framework and the analytical framework. The computational framework detects chick vocalisations, extracts acoustic features using signal processing, and identifies call clusters. 
The analytical framework is built upon the resulting call-level dataset, applying multivariate analyses and linear mixed-effects modelling to quantify differences in vocal production and acoustic features between VPA-exposed chicks and controls. Analyses were performed in Python (version 3.9) for the computational framework and R (version 4.3) for the analytical framework.

\subsection{Datasets}

We used two distinct datasets. We established the computational framework using the development dataset and then applied it to the experimental dataset, derived from a previous study \cite{lorenzi2019embryonic}. 
In both cases, audio recordings were extracted from videos acquired using a Microsoft LifeCam, a general-purpose webcam with a built-in omnidirectional super-wideband microphone (frequency response 100~Hz--18~kHz), rather than specialised acoustic equipment. 
Audio files were therefore converted to 16-bit WAV format (44.1 kHz) using Audacity~\cite{audacity2017}. Stereo signals were equalised and merged into mono-channel recordings to ensure balanced input contribution.
The duration of the recordings matched the duration of the original experiments.

\subsubsection{Development dataset}

This dataset consists of 31 mono-channel audio recordings (WAV format, 44.1 kHz, 16-bit) of visually naïve chicks (19 males, 12 females) tested between 12 and 36 hours after hatching. Each chick was individually recorded for $\sim$10 minutes during free exploration in a rectangular wooden arena (90×60 cm). Vocalisation onsets and offsets were manually annotated by three trained experts using Sonic Visualiser \cite{cannam2006sonic}, after a training in which they reached a minimum inter-rater reliability of 90\% in a subset of examples.

For the onset and offset detection task, the development dataset was divided into three subsets, balanced for age and sex: a training set (19 files: 10 males, 9 females; 13,261 calls), a validation set (6 files: 5 males, 1 female; 3,471 calls), and a testing set (6 files: 4 males, 2 females; 3,486 calls). The total duration of the analysed recordings was 5 hours and 26 minutes.
For feature extraction and clustering analysis, we selected a subset (12 audio files: 7 males, 5 females; 5,924 calls) with low background noise from the training, validation, and testing datasets.

\subsubsection{Experimental dataset}

This dataset consists of 46 mono-channel audio recordings (WAV format, 44.1 kHz, 16-bit) of visually naïve chicks tested within 12 hours after hatching. Each chick was individually recorded for 6 minutes during free exploration in a rectangular wooden arena (85×30 cm). Two objects were projected at the opposite ends of the arena: one object showed changes in speed, while the other maintained a constant speed. The dataset includes a VPA-treated group (27 chicks: 16 males, 11 females; 14,521 calls) and a control vehicle-injected group (19 chicks: 8 males, 11 females; 9,028 calls). Detailed experimental procedures are described in \cite{sgado2018embryonic,lorenzi2019embryonic, adiletta2021spontaneous, adiletta2022embryonic}.
The total duration of the analysed recordings was 4 hours and 18 minutes.

\subsection{Computational framework}

\subsubsection{Pre-processing, onset and offset detection}

As an initial step in preprocessing, audio recordings underwent max loudness normalisation to enhance the precision of call detection algorithms and ensure comparability across experiments and sessions.
For call onset and offset detection, we used the best algorithms identified in a previous study on the development dataset \cite{torrisi2024exploratory}. For onset detection, we used the High Frequency Content (HFC) algorithm \cite{muller2015fundamentals}, which detects frequency changes within the signal. The algorithm was implemented using the \texttt{madmom} library \cite{bock2016madmom}.
 
For offset detection, we implemented a method based on first-order differencing of the energy signal, followed by local minimum detection within a predefined time window based on the maximum expected duration of chick calls. Both methods outperformed alternative approaches tested (weighted F1-measure = 0.85 and 0.72, respectively, on the development dataset; see Supplementary Material~\ref{sec:extended_method_onset_offset}).

To assess the generalisability of the results obtained with our automated approach, we conducted a comprehensive benchmarking evaluation on the annotated experimental dataset. The HFC algorithm achieved an excellent onset detection performance (weighted F1-measure = 0.933, precision = 0.910, recall = 0.961). For the offset detection, the first-order difference method also performed well (weighted F1-measure = 0.885, precision = 0.885, recall = 0.885). Methodological details are reported in the Supplementary Material~\ref{sec:s1_onset_offset}.

\subsubsection{Feature extraction}

Before feature extraction, a bandpass filter (BPF) was applied to reduce background noise and focus on the frequency bands relevant to chick vocalisations, specifically the fundamental frequency and the first two harmonics. The frequency range of the filter was set to 2000--12,600 Hz for the development dataset and to 2000--15,000 Hz for the experimental dataset, as preliminary visual inspection indicated that the chicks' formants reached slightly higher frequencies in the experimental recordings; therefore, the upper bound was set to preserve harmonic content.

For each segmented call, we extracted different signal components using established methods (for details see Supplementary Material \ref{sec:extended_method_feature_extraction}): Fundamental Frequency (F0) via PYIN algorithm \cite{mauch2014pyin}, first and second harmonics (F1, F2) and their magnitudes via FFT \cite{muller2015fundamentals}, Root Mean Square (RMS) \cite{Panagiotakis}, Spectral Centroid \cite{muller2015fundamentals}, and signal Envelope via Hilbert transform \cite{virtanen2018computational}.

From these signal-level features, we characterised each call by calculating 20 one-dimensional statistical descriptors, grouped by domain: temporal (Call Duration, Attack Time, Envelope Slope), frequency (F0 Mean, F0 Standard Deviation, F0 Kurtosis, F0 Skewness, F0 Bandwidth, F0 Slope, F0 1st Order Difference Mean, F0 Mag Mean, F1 Mag Mean, F2 Mag Mean, F1/F0 Ratio Mean, F2/F0 Ratio Mean, Spectral Centroid Mean and Spectral Centroid Standard Deviation), and energy (RMS Mean, RMS Standard Deviation, Attack Magnitude). 
Each descriptor is a within-call statistic, computed from the frame-level values of the corresponding signal feature. Feature extraction was performed using the \texttt{librosa} \cite{mcfee2015librosa} and \texttt{SciPy} \cite{virtanen2020scipy} Python libraries.
Detailed feature descriptions and equations to derive them are provided in Table~\ref{tab:features_and_equations}. The features quantify complementary aspects of each call: for example, RMS Mean measures overall intensity; F0 variability descriptors measure the modulation of pitch across the call; and the Spectral Centroid, the amplitude-weighted mean of the frequency spectrum, reflects the call's predominant frequency content, whose perceptual correlate is brightness \cite{marozeau2007effect}.

All these features are biologically relevant for the study of chicks' vocalisations \cite{collias1953spectrographic, collins2024sound}. More details are available in Supplementary Section~\ref{features_interpretation}. Features were z-score normalised prior to clustering, ensuring they were on a comparable scale for these analyses.

\subsubsection{Clustering analysis and UMAP}
\label{sec:clustering_methodology}

For the initial exploratory analysis of chicks’ vocal repertoire, we tested hard clustering methods (K-means, Hierarchical Agglomerative Clustering, DBSCAN) and soft clustering methods (Fuzzy C-Means, Gaussian Mixture Models) to capture discrete categories or continuous distributions \cite{fischer2017structural,zadeh2008there,wadewitz2015characterizing}. 

To identify the optimal clustering structure for the development dataset, we performed a grid search, testing solutions from 1 to 10 clusters for each algorithm. 
Three main metrics guided the selection of the number of clusters: (i) the Silhouette Score \cite{rokach2023machine}, which quantifies how similar each call is to its own cluster compared to others; (ii) the Within-Cluster Sum of Squares (WCSS) \cite{rokach2023machine} used with the Elbow method to identify the point beyond which adding more clusters does not improve compactness; and (iii) the Calinski-Harabasz Index (CHI) \cite{rokach2023machine}, which evaluates the ratio of between- to within-cluster dispersion. 
For soft clustering methods, we used additional metrics: Fuzzy Partition Coefficient (FPC) \cite{rokach2023machine} for Fuzzy C-Means, and AIC and BIC \cite{rokach2023machine} for Gaussian Mixture Models (see details in Supplementary Material \ref{sec:clustering_supplementary}). The single-cluster solution was evaluated through WCSS and, for Gaussian Mixture Models, through AIC and BIC, as the Silhouette Score and CHI require at least two clusters (see Supplementary Section~\ref{sec:clustering_results_development_dataset}).
We considered these metrics jointly, as each captures a different aspect of cluster structure \cite{goffinet2021low}. In graded vocal repertoires, silhouette values can be low even in the presence of meaningful structures \cite{fischer2017structural, sainburg2020finding}, supporting the use of multiple metrics. Clustering methods and evaluation metrics were implemented using the \texttt{scikit-learn} library \cite{pedregosa2011scikit}, while for Fuzzy C-Means we used the \texttt{scikit-fuzzy} library (0.4.2).

After identifying the best cluster division, we inspected calls at varying distances from the cluster centroids to assess within-cluster homogeneity and identify potential outliers, using a percentile-based sampling method: calls within each cluster were ranked by Euclidean distance from centroids, which represent the mean acoustic profile of each cluster. Across methods, Hierarchical Agglomerative Clustering (HAC) produced the most internally consistent clusters, with the fewest outliers; we therefore report the results of this method. The same grid search approach was then applied to the experimental dataset using HAC and evaluated through the Silhouette Score, CHI, and WCSS.
Hierarchical clustering structure for both datasets was visualised through dendrograms, where branch height represents the Ward distance at which clusters merge.

Finally, we projected all calls into a two-dimensional space using Uniform Manifold Approximation and Projection (UMAP) \cite{goffinet2021low, mcinnes2018umap}. The visualisation in the embedding space allowed us to inspect the distribution of calls, the separation of clusters, and the overlap between the experimental groups. Details are reported in Supplementary Section~\ref{sec:umap_supplementary}. 
 
\subsection{Analytical framework for biological comparisons}

To assess whether embryonic VPA exposure alters chick vocalisations, we applied an analytical framework to the output of the computational framework. We progressed from call production analysis to fine-grained acoustic characterisation of vocalisation clusters. To analyse the effect of VPA exposure on the total number of calls, we fitted a Linear Mixed-Effects Model (LMM) with time bin (1-6 minute), experimental condition (control \textit{versus} VPA), call cluster membership, and their interactions as fixed effects. Individual chick identity was included as a random intercept to account for repeated measures. Models were fitted using the \texttt{lme4} \cite{bates2015fitting} and \texttt{lmerTest} \cite{kuznetsova2017lmertest} R packages. Likelihood ratio tests ($\alpha = 0.05$), supported by AIC values, were used to identify the minimum adequate model via progressive simplifications \cite{harrison2018brief}.

To investigate the effect of VPA exposure on acoustic features, we reduced multicollinearity by selecting one feature from each highly correlated pair (Pearson |r| ≥ 0.8) based on discriminative power (Cohen's d) or biological interpretability (see Supplementary material, Section \ref{sec:multicollinearity_supplementary} for details). Each feature was analysed using the same LMM approach described above.
We estimated effect sizes as partial eta-squared ($\eta^2_p$) and applied False Discovery Rate (FDR) correction across all likelihood ratio tests to control for multiple testing (see Supplementary Section~\ref{sec:lmm_supplementary}).

To understand whether VPA produced cluster-specific effects, we conducted Tukey-adjusted post hoc comparisons using the \texttt{emmeans} package \cite{lenth2024emmeans}. Results were organised by acoustic domain (temporal, energy, and frequency domains) to facilitate biological interpretation.

\section{Results}

\subsection{Computational results}

\subsubsection{Clustering results in the development dataset}

\begin{figure}[h]
 \centering
 \includegraphics[width=1.05\linewidth]{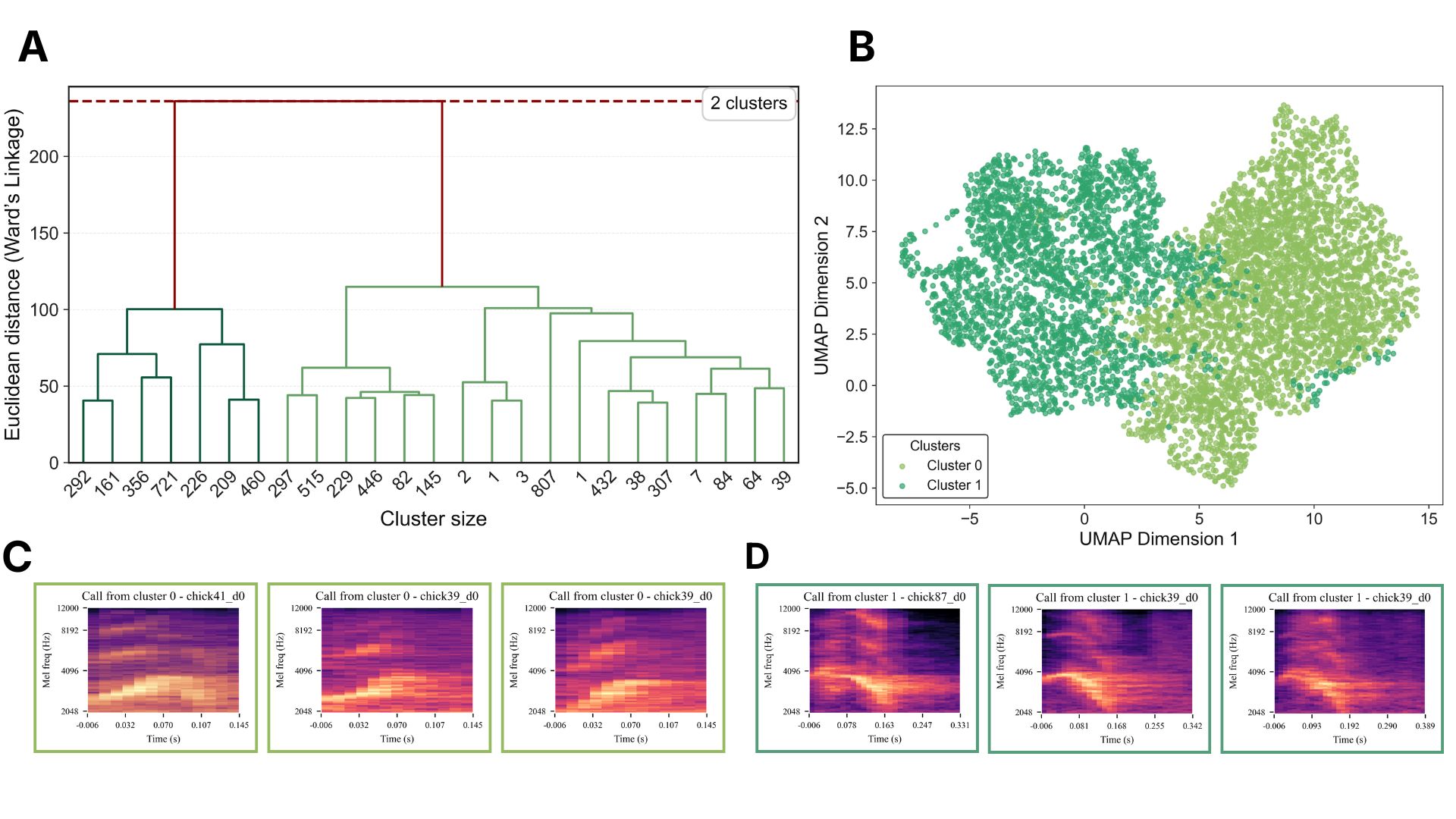}
\caption{A) Dendrogram of the development dataset showing the clustering structure and optimal cut points. Within the main clusters, we observed further branching; B) UMAP projection divided into \( K = 2 \) clusters using HAC; C) Spectrograms of representative calls extracted from cluster 0; D) Spectrograms of representative calls extracted from cluster 1.}
 \label{fig:clustering_umap_dev_dataset}
\end{figure}

HAC identified two optimal clusters in chicks' vocal repertoire (Figure~\ref{fig:clustering_umap_dev_dataset}A), confirmed by Silhouette at \( K = 2 \) (0.262) and CHI (1820.717). UMAP visualisation (Figure~\ref{fig:clustering_umap_dev_dataset} B) shows that chicks' vocalisations are distributed along a continuum, rather than occupying two distinct embedding spaces.

The two clusters showed distinct acoustic profiles, defined by their mean feature values: cluster 0 comprised soft calls with shorter duration, lower pitch, and lower attack time, whereas cluster 1 comprised calls with higher energy, longer duration, and higher attack time. The mean feature values for each cluster are reported in Supplementary Table~\ref{tab:combined_cluster_characteristics}. Qualitative inspection of representative calls afterwards confirmed these profiles. 
Spectrograms of representative calls for each cluster are shown in Figure~\ref{fig:clustering_umap_dev_dataset} C and D, for illustrative purposes. Representative calls are derived from the 5th percentile of the Euclidean distance-to-centroid distribution, representing the most central and characteristic vocalisations for each cluster.

\subsubsection{Clustering analysis results in the experimental dataset (VPA \textit{versus} control)}

In the experimental dataset, we applied the HAC method independently to the VPA and control groups. For the VPA-treated group, HAC showed an optimal clustering division at \( K = 2 \), as indicated by the peak of Silhouette Score (0.425) and CHI (1724.68; see Supplementary Figure~\ref{fig:VPA_silhouette_chi}). The Elbow method applied to the WCSS identified \( K = 4 \) (WCSS = 219535.80; see Supplementary Figure~\ref{fig:VPA_wcss}) as optimal division, suggesting a possible substructure in the data. However, the relatively low Silhouette Score at \( K = 4 \) (0.125) suggested poor cluster separation, and therefore an implausible division.

\begin{figure}[h]
 \centering
 \includegraphics[width=1.01\linewidth]{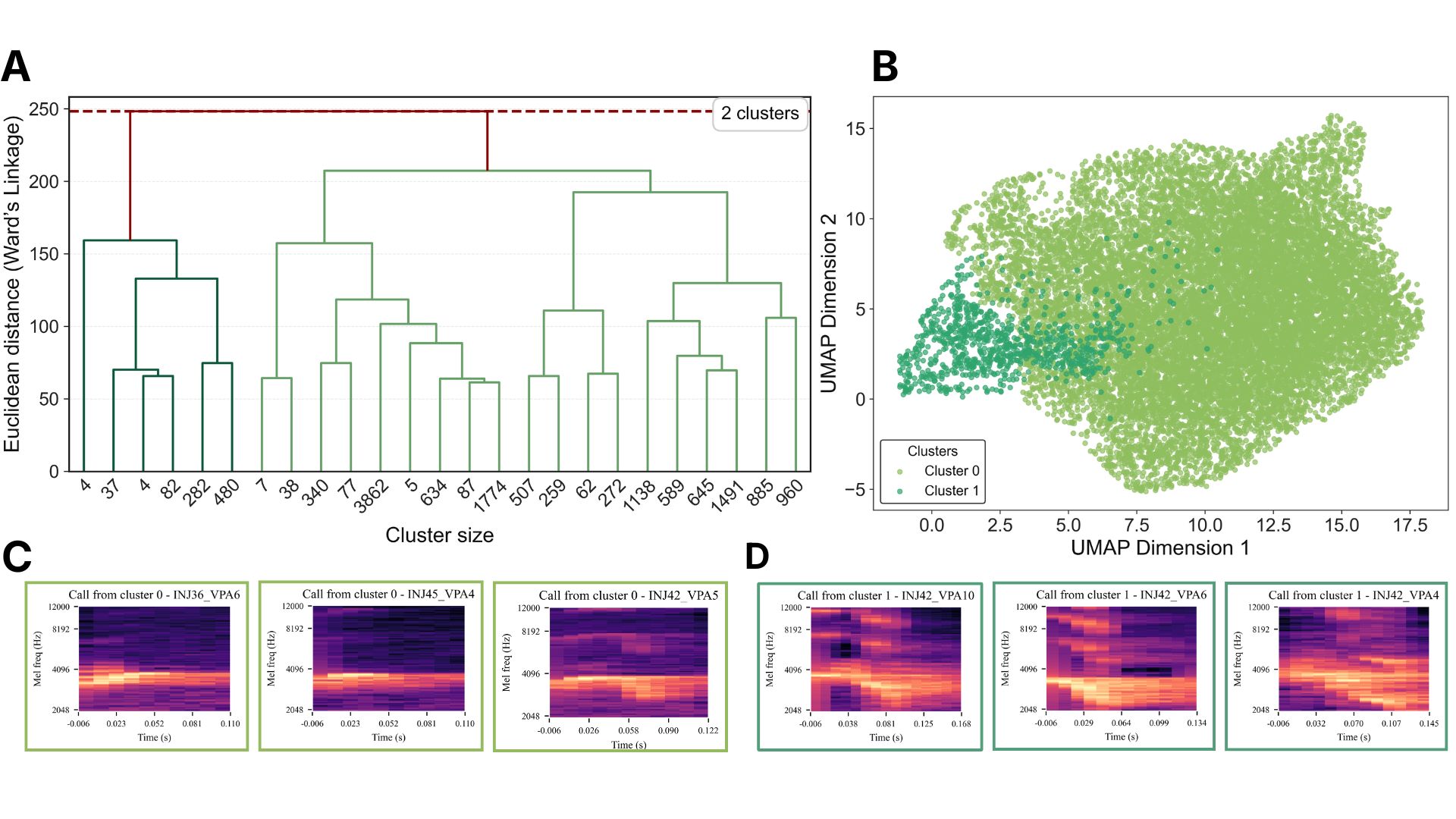}
\caption{A) Dendrogram for VPA-treated chicks showing the clustering structure and optimal cut points. Within the main clusters, we observed further branching; B) UMAP projection for VPA-treated chicks with \( K = 2 \) clusters; C) Spectrograms of representative calls extracted from cluster 0; D) Spectrograms of representative calls extracted from cluster 1.}
\label{fig:clustering_umap_VPA_dataset}
\end{figure}

The dendrogram structure supports the quantitative findings, showing clear branching patterns that align with the \( K = 2 \) solution (see Figure~\ref{fig:clustering_umap_VPA_dataset}A). UMAP projection at \( K = 2 \) for VPA-exposed chicks (Figure~\ref{fig:clustering_umap_VPA_dataset}B) shows cluster overlap, and a higher number of calls assigned to cluster 0.

\begin{figure}[h]
 \centering
 \includegraphics[width=1.01\linewidth]{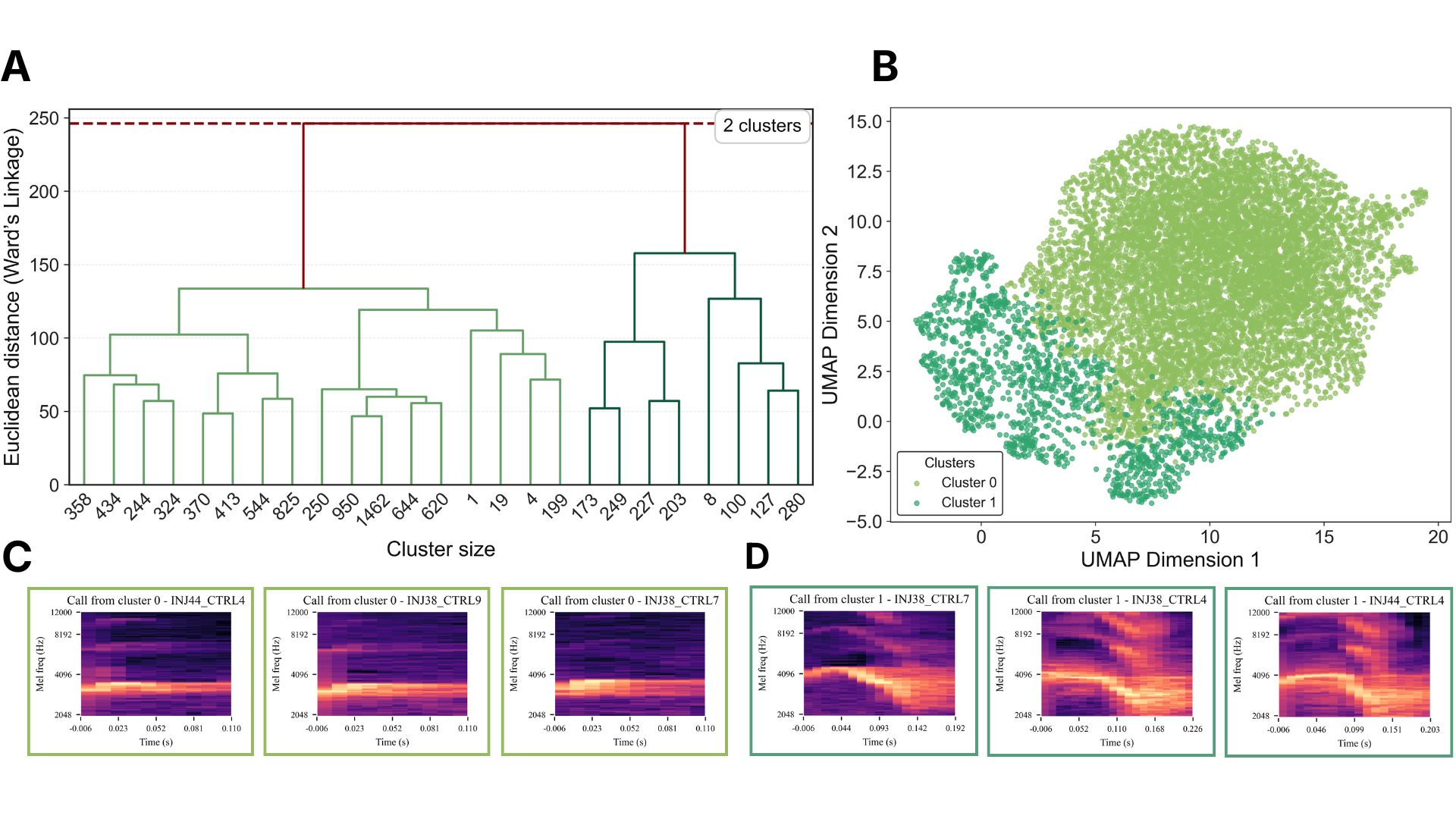}
\caption{A) Dendrogram for control chicks showing the clustering structure and optimal cut points. Within the main clusters, we observed further branching; B) UMAP projection for control chicks with \( K = 2 \) clusters; C) Spectrograms of representative calls extracted from cluster 0; D) Spectrograms of representative calls extracted from cluster 1.}
\label{fig:clustering_umap_ctrl_dataset}
\end{figure}

Similarly, the control group showed optimal clustering at \( K = 2 \) (Silhouette Score: 0.393; CHI: 1820.28; see Supplementary Figure~\ref{fig:control_silhouette_chi}), with \( K = 3 \) (Silhouette Score: 0.347) also providing a reasonable balance between separation and model complexity. The Elbow method identified \(K = 3\) (WCSS = 137809.06; see Supplementary Figure~\ref{fig:control_wcss}) as optimal cluster division. However, the corresponding Silhouette Score (0.347) remained lower than at \(K = 2\). The dendrogram for the control chicks showing the cluster division into 2 groups can be observed in Figure~\ref{fig:clustering_umap_ctrl_dataset}A. The UMAP representation (Figure~\ref{fig:clustering_umap_ctrl_dataset}B) shows that the control group displayed more clearly delineated clusters compared to the VPA group. A higher number of calls is assigned to cluster 0, especially in the VPA group.

Representative calls nearest to cluster centroids are shown in Figures~\ref{fig:clustering_umap_VPA_dataset}C and D, and \ref{fig:clustering_umap_ctrl_dataset}C and D.

To ensure cluster comparability between groups, we verified correspondence by examining their acoustic profile, i.e., the mean feature values of each cluster from each group (as reported in Supplementary Table~\ref{tab:combined_cluster_characteristics}). We then qualitatively inspected representative spectrograms. Calls assigned to cluster 0 in the VPA group closely resemble those in cluster 0 of the control group. Similarly, the calls in cluster 1 of the VPA group are comparable to those in cluster 1 of the control group. Calls in cluster 0 appear shorter in duration, characterised by lower frequency and energy, and lack the second harmonic in both groups. These qualitative observations confirmed that the clustering procedure captured consistent vocal categories across experimental groups, with comparable acoustic profiles emerging for analogous clusters. However, subtle variations in temporal, frequency, and energy domain acoustic features may reflect the underlying effects of VPA treatment on vocal production. The quantitative distribution of calls across clusters confirmed these qualitative observations and revealed a pronounced imbalance in both groups. In the control group, 7,661 calls were assigned to cluster 0 and 1,367 to cluster 1. In the VPA group, the imbalance was more pronounced, with 13,632 calls in cluster 0 and only 889 calls in cluster 1.

\subsection{Analytical results}

\subsubsection{Number and rate of calls in the experimental dataset} 

\begin{figure}[ht]
\centering
\includegraphics[width=0.65\textwidth]{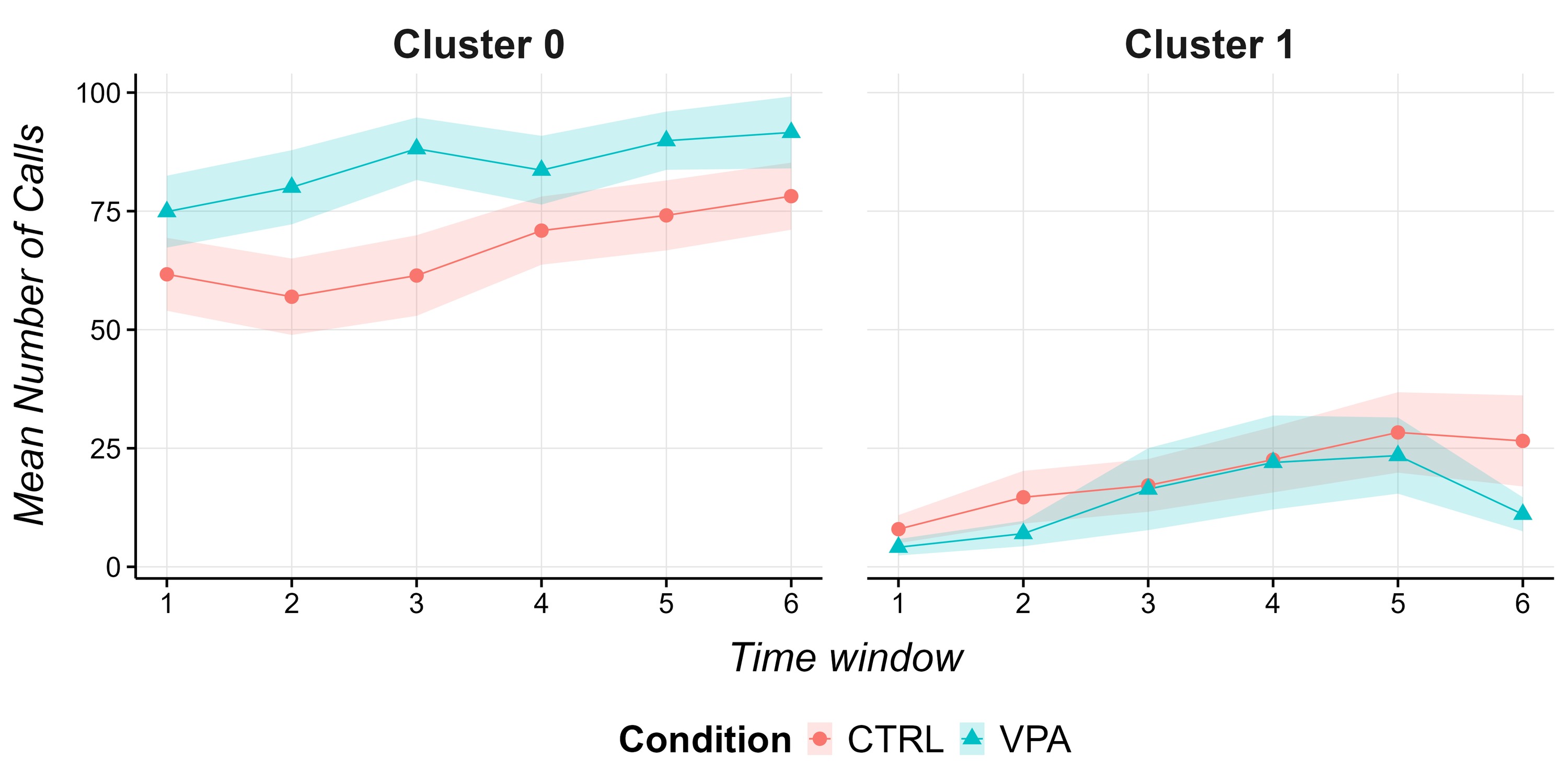}
\caption{Mean number of calls across 6 time bins by condition and cluster membership. Mean ± SEM.}
\label{fig:calls_by_cluster_condition}
\end{figure}

For the number of calls, the minimum adequate model included the interaction between Condition and Cluster membership and the main effect of Time bin (see Supplementary Table~\ref{tab:lmm_calls}), indicating VPA exposure affects the two call types differently. The number of calls increased over time, with more calls in the last two time bins than in the first two (all p < 0.05; Figure~\ref{fig:calls_by_cluster_condition}).

Both groups produced significantly more cluster 0 than cluster 1 calls (control: 67.2 vs 19.2 calls per time bin, p < 0.001; VPA: 84.4 vs 8.3 calls per time bin, p < 0.001). VPA exposure increased cluster 0 calls (+17.2 calls per time bin, p = 0.012), whilst cluster 1 calls did not differ significantly between conditions (−10.9 calls per time bin, p = 0.152).

\subsubsection{Feature analysis}
\label{sec:lmm_main_section}

After excluding multicollinear features, the analysis included, for the temporal domain: Call Duration and Envelope Slope; for the energy domain: RMS Mean; for the frequency domain: F0 Standard Deviation, F0 Skewness, F0 Kurtosis, F0 1st Order Difference, F0 Slope, F0 Magnitude Mean, F1 Magnitude Mean, F2 Magnitude Mean, F1/F0 Ratio Mean, F2/F0 Ratio Mean, Spectral Centroid Mean and Spectral Centroid Standard Deviation. Full results are reported in Supplementary Table~\ref{tab:lmm_complete}.

\paragraph*{Temporal domain features.}

For the Call Duration, the minimum adequate model included the interaction between Condition, Cluster membership and Time bin (see Supplementary Table~\ref{tab:lmm_complete}, Figure~\ref{fig:temporal_features}). For cluster 0 calls, we observed no differences between conditions, whilst for cluster 1 calls, the Duration was longer for control chicks compared to VPA chicks, and this difference became stronger over time.
Results for Envelope Slope are reported in the Supplementary Material (Section~\ref{sec:lmm_supplementary}, Figure~\ref{fig:attack_slope_complete}).

\begin{figure}[htbp]
 \centering
  \includegraphics[width=0.65\textwidth]{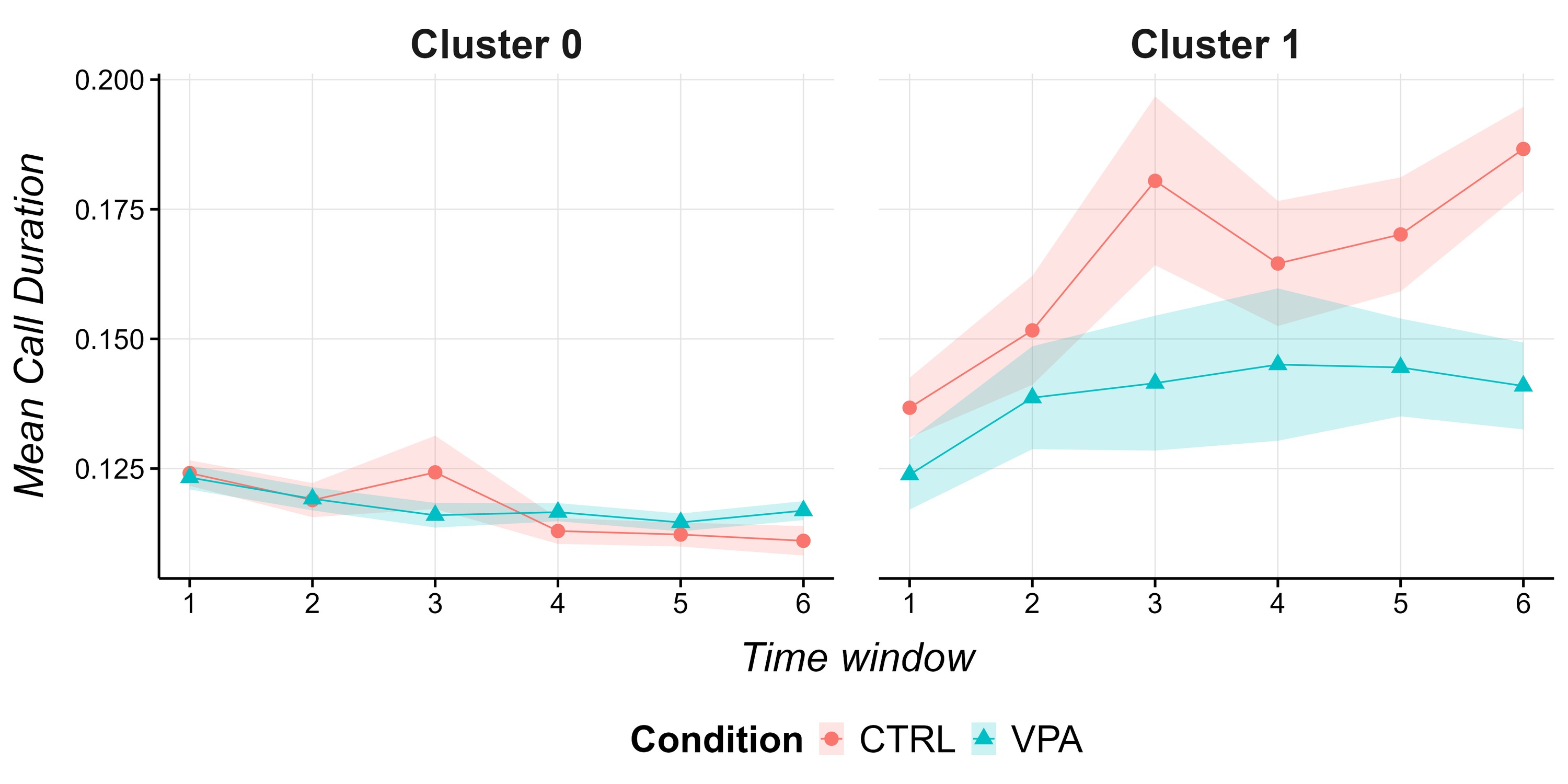}
  \caption{Call Duration across 6 time bins, plotted by condition and cluster (2-cluster solution). Mean ± SEM.}
  \label{fig:temporal_features}
\end{figure}

\paragraph*{Energy features.}

\begin{figure}[htbp]
 \centering
 \includegraphics[width=0.65\textwidth]{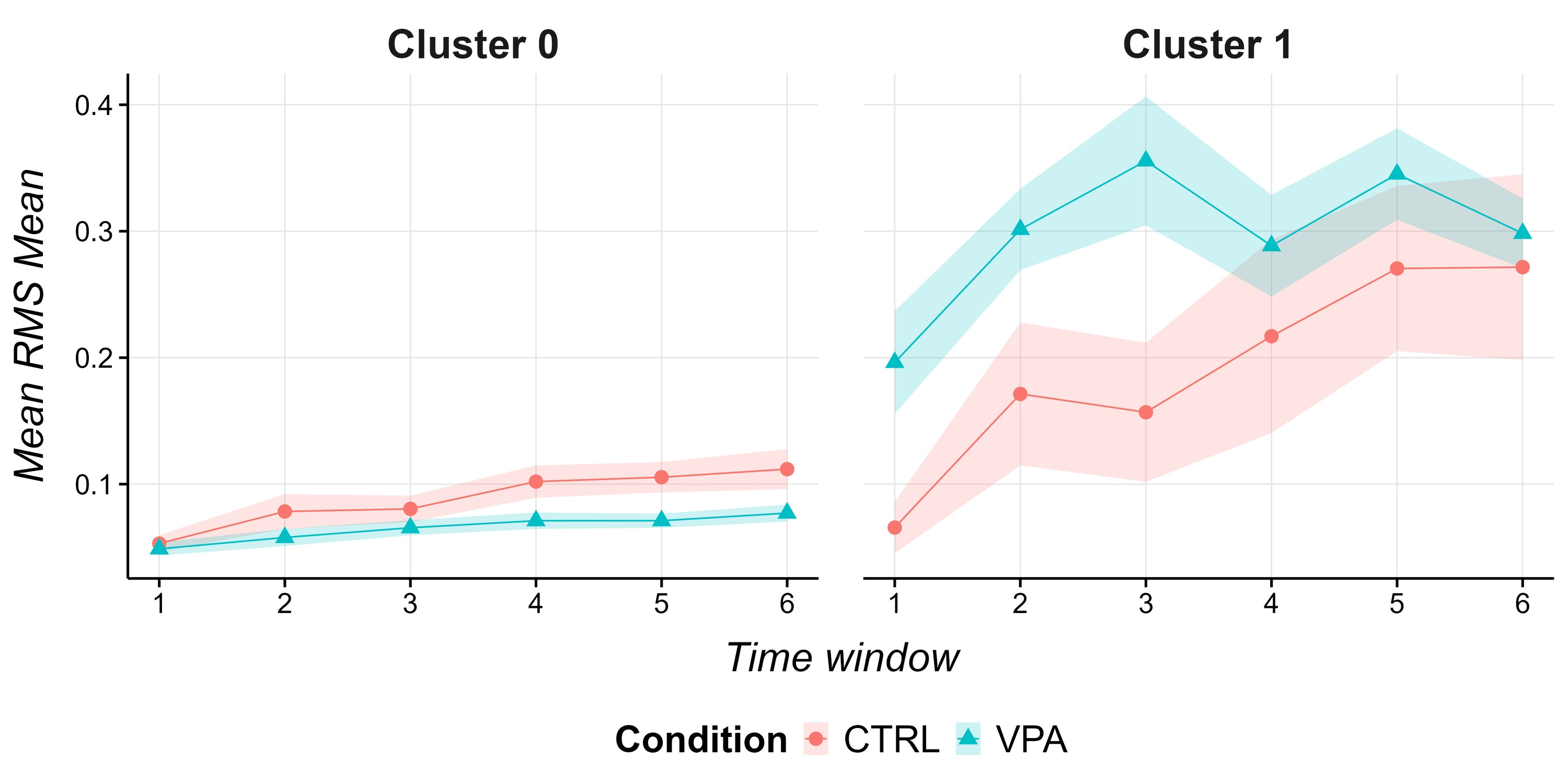}
 \caption{RMS Mean across 6 time bins, plotted by condition and cluster (2-cluster solution). Mean ± SEM.}
 \label{fig:energy_features}
\end{figure}

For the RMS mean, the minimum adequate model included the interaction between Condition, Cluster membership and Time bin (see Supplementary Table~\ref{tab:lmm_complete}). While cluster 0 calls showed lower RMS mean values in VPA compared to control chicks in later bins, cluster 1 calls exhibited the opposite pattern. These condition-dependent differences varied over time, as shown in Figure~\ref{fig:energy_features}.

The additional energy features (RMS Standard Deviation and Attack Magnitude), excluded from the main analysis due to their high correlation with RMS Mean, are presented in the Supplementary material section~\ref{sec:lmm_supplementary} (Supplementary Table~\ref{tab:lmm_complete} and Figure~\ref{fig:Energy_related_features_suppl}).

\paragraph*{Frequency domain features.} 

Similar patterns emerged for fundamental frequency variability features (F0 Standard Deviation, F0 Kurtosis, F0 1st Order Difference, F0 Slope, and F0 Skewness), with the minimum adequate model including the interaction between Condition, Cluster membership, and Time bin (Supplementary Table~\ref{tab:lmm_complete}).

\begin{figure}[htbp]
 \centering
 \includegraphics[width=\textwidth]{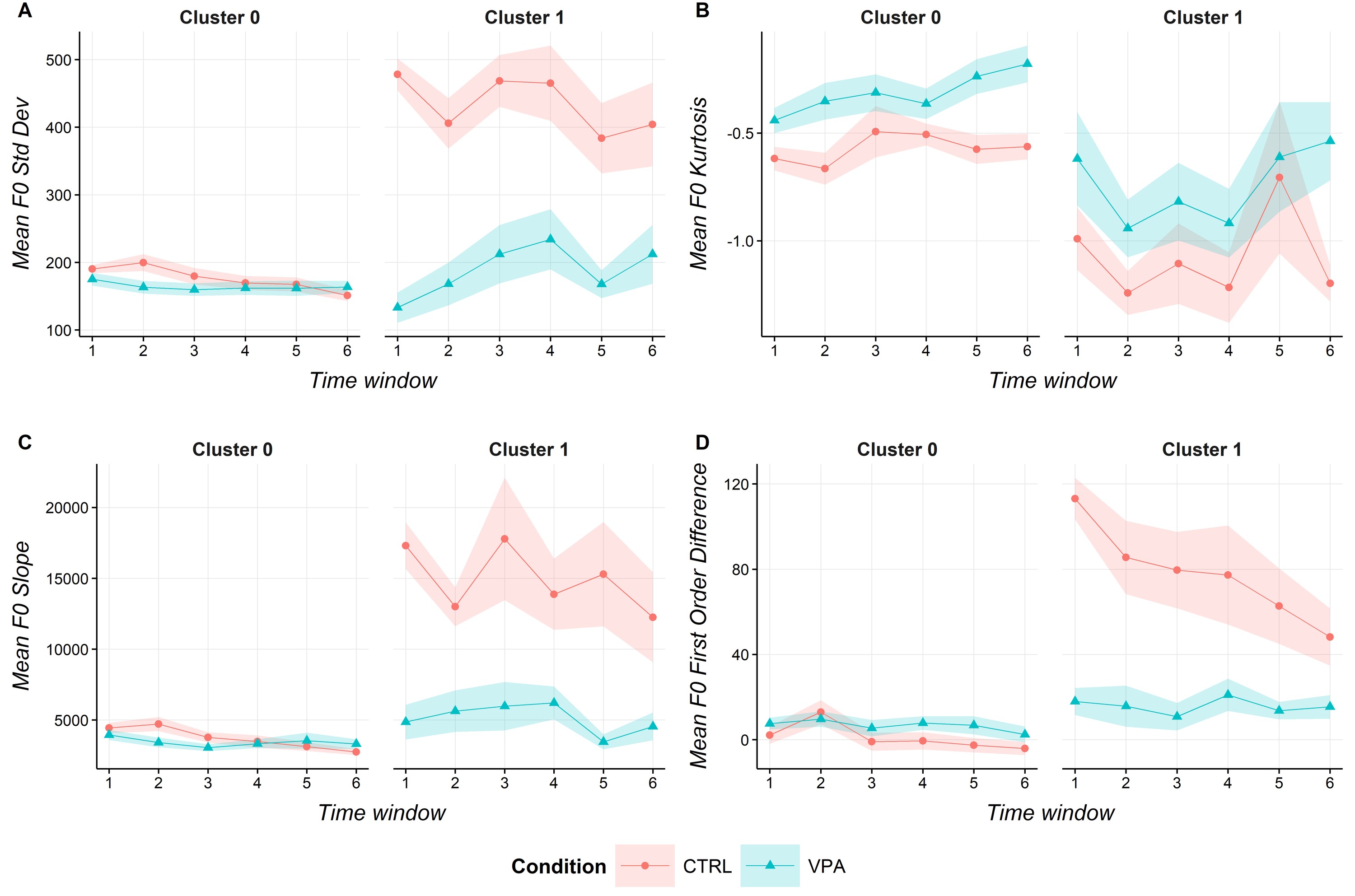}
 \caption{Frequency-domain features across 6 time bins by condition and cluster. 
 (A) F0 Standard Deviation, (B) F0 Kurtosis, (C) F0 Slope, (D) F0 1st Order Difference. Mean ± SEM.}
 \label{fig:frequency_features_1}
\end{figure}

For F0 Standard Deviation, VPA exposure reduced F0 variability in cluster 1 calls (Figure~\ref{fig:frequency_features_1}A), suggesting flattened pitch dynamics over the session, while no significant difference was found for cluster 0 calls between conditions. %contact calls are flatter for VPA and become flatter over six minutes ( bin significant),
For F0 Kurtosis, control chicks showed lower F0 Kurtosis in cluster 1 versus cluster 0 calls, indicating flatter pitch distribution, while VPA-treated chicks showed reduced cluster differences, making cluster 1 calls more similar to cluster 0 calls in pitch distribution (Figure~\ref{fig:frequency_features_1}B).
For F0 Slope, while cluster 0 calls showed no condition differences, cluster 1 calls had reduced F0 slope values in VPA compared to control chicks. This attenuation was evident in the early temporal bins, effectively diminishing the usual separation in pitch change rate between the two call types (see Figure~\ref{fig:frequency_features_1}C).
Similarly, for the F0 first-order difference, patterns broadly mirrored those observed for the F0 Slope. Cluster 0 calls showed no differences between conditions, while cluster 1 calls exhibited reduced F0 first-order difference variation (narrowing the typical divergence in pitch modulation between clusters) in VPA compared to control chicks. This difference was most pronounced in early time bins (see Figure~\ref{fig:frequency_features_1}D).
For F0 Skewness, we observed similar patterns of interaction. However, these did not remain significant after FDR correction, and effect sizes were smaller than those of the other frequency variability features (Supplementary material ~\ref{sec:lmm_supplementary} and Figure~\ref{fig:frequency_features_suppl}). Overall, VPA embryonic exposure altered the frequency modulation of cluster 1 calls.

\begin{figure}[ht]
 \centering
 \includegraphics[width=\textwidth]{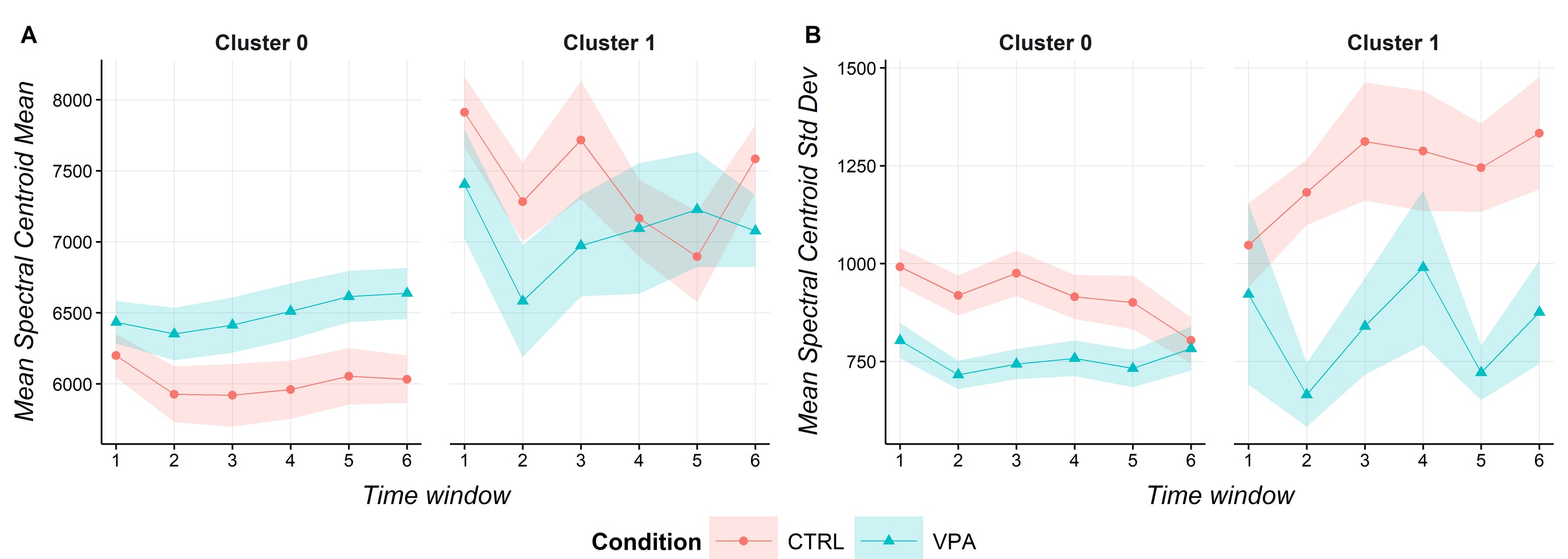}
\caption{Spectral features across 6 time bins by condition and cluster. 
(A) Spectral Centroid Mean, (B) Spectral Centroid Standard Deviation. 
Mean ± SEM.}
 \label{fig:spectral_features}
\end{figure}

Spectral features showed condition-dependent changes and varied between clusters. For the Spectral Centroid Mean, the minimum adequate model included two-way interactions, as the three-way interaction was not significant (see Supplementary Table~\ref{tab:lmm_complete}). Cluster 0 calls showed higher spectral centroid values in VPA compared to control chicks across all time bins, while cluster 1 calls exhibited a complex temporal pattern with VPA effects varying relative to controls across the session (see Figure~\ref{fig:spectral_features}A). For the Spectral Centroid Standard Deviation, the minimum adequate model included the interaction between Condition, Cluster membership and Time bin (see Supplementary Table~\ref{tab:lmm_complete}). Both clusters showed consistently reduced spectral variability under VPA conditions compared to controls, with cluster 1 exhibiting larger decreases than cluster 0, indicating that VPA exposure leads to more stereotyped spectral characteristics for both call types (see Figure~\ref{fig:spectral_features}B).

Additional frequency features relating to the magnitude of the fundamental frequency and its formants (F0 Magnitude Mean, F1 Magnitude Mean, F2 Magnitude Mean, F1/F0 Ratio Mean, F2/F0 Ratio Mean) are presented in Supplementary material in section~\ref{sec:lmm_supplementary} (Supplementary Table~\ref{tab:lmm_complete} and Figures~\ref{fig:frequency_features_suppl} and~\ref{fig:freq_features_formants}).

These features revealed distinctive patterns based on their acoustic properties. Magnitude features (F0 Magnitude Mean, F1 Magnitude Mean) demonstrated similar temporal inversion patterns. VPA exposure caused progressive declines specifically in cluster 1 calls across the recording session while control values increased, resulting in marked temporal inversions where VPA-treated chicks showed reduced magnitude in later time bins compared to controls (Figures~\ref{fig:frequency_features_suppl} and~\ref{fig:freq_features_formants}).
In contrast, ratio features (F1/F0 Ratio Mean and F2/F0 Ratio Mean) showed more stable temporal dynamics without marked inversions. VPA exposure was associated with lower ratio values, indicating weaker harmonics compared to the fundamental frequency. For the F1/F0 Ratio Mean, this effect was significant in cluster 0 calls, while for the F2/F0 Ratio Mean the effect of condition did not reach significance.
These patterns suggest that VPA exposure affects both signal magnitude and spectral balance (Figure~\ref{fig:freq_features_formants}).

\section{Discussion}

We introduced a computational framework that integrates call detection, multi-feature extraction, and unsupervised learning for the analysis of chicks' vocal repertoire. We first characterised the natural structure of the early vocal repertoire by applying our approach to a dataset of newly hatched untreated chicks. Clustering analyses revealed vocalisations distributed along a continuum between two main categories, rather than in discrete ones. Whereas previous studies have described chick vocalisations as a set of discrete call types \cite{marx2001vocalisation, thomas2023using}, our unsupervised analysis reveals a graded organisation of the early vocal repertoire that has not been reported before. Calls assigned to cluster 0 were shorter, lower in energy, and showed ascending frequency modulation, while calls assigned to cluster 1 were longer, louder, and with higher fundamental frequency values. Subcategories of calls are present within these two clusters, with more strongly supported subdivisions in cluster 0. Overall, cluster 0 appears to include positive affective state calls such as previously described pleasure calls \cite{collias1953spectrographic}, whereas cluster 1 calls appear to include high arousal and contact calls, as described by \cite{collias1953spectrographic} and \cite{marx2001vocalisation}. The exact correspondence of these clusters (and sub-clusters) to specific communicative functions and affective states should be further studied across different experimental settings.

The graded organisation of the repertoire also highlights the value of an unsupervised approach. Along a continuum, the boundaries between call types are difficult to define a priori, whereas our framework derives them from the data and still captured biologically relevant differences between VPA-exposed and control chicks.

We then applied the same pipeline to an independent dataset of chick calls \cite{lorenzi2019embryonic} to assess whether prenatal VPA exposure affects early vocal production. The analyses confirmed the two-cluster continuum structure, with acoustic profiles resembling those in the development dataset: calls in cluster 0 were shorter and lower in energy, while those assigned to cluster 1 were longer and louder, with higher frequency values. This consistency supported the robustness of our approach and enabled the assessment of systematic differences between control and VPA-exposed chicks.

% Scientific findings as extension of previous work on the technical side
Our framework addresses several methodological limitations of previous chick vocalisation studies.
We used signal processing techniques for detecting the onset and offset of individual calls, achieving high performance across all metrics (see Table~\ref{tab:onset_offset_performance}). This approach eliminates the need for manual annotation, on which previous studies relied \cite{marx2001vocalisation}, while avoiding potential temporal boundary artefacts introduced by sliding-window methods that may capture incomplete calls \cite{grzywalski2025broiler}.
Each segmented call is then characterised by 20 acoustic features across temporal, frequency, and energy domains, capturing variability beyond traditional prototypical call classifications (see Table~\ref{tab:literature_computational_comparison} for a literature comparison), while ensuring interpretability, a critical balance often lost in black-box approaches.
By avoiding taxonomic priors \cite{collias1953spectrographic}, our approach derives repertoire structure directly from the data, rather than imposing predefined categories as in supervised methods \cite{thomas2023using}. This reduces classifier biases and ensures reproducible, automated processing of large datasets. The resulting broader acoustic space illustrates the value of data-driven methods for uncovering the full spectrum of natural vocal variation.

% Scientific findings as extension of previous work for the bio-side 
Applying this framework to explore the early vocal repertoire of VPA-exposed chicks, we extend previous work on avian and mammalian models by identifying call-type-specific feature differences \cite{matsushima2024domestic,tewelde2025prenatal,nicolini2018valproic, csillag2022avian}. VPA-exposed chicks produced proportionally more cluster 0 calls and fewer cluster 1 calls, the latter typically linked to contact calls for conspecifics. This parallels findings in marmosets \cite{mimura2026altered}, where VPA exposure altered call type distributions and disrupted parent-infant vocal interactions, suggesting that VPA affects specific call types rather than uniformly disrupting vocal production across species \cite{uesaka2023automatic, nicolini2018valproic}.

Feature analysis showed systematic effects of VPA exposure with significant interactions between condition and call type across temporal, energy and frequency domains. In the temporal domain, VPA-exposed chicks showed reduced cluster 1 call duration. These findings align with rodent studies demonstrating that prenatal VPA exposure reduces call duration across developmental stages \cite{gzielo2020valproic, moldrich2013inhibition}.

In the energy domain, RMS Mean showed opposite trends across clusters. VPA-exposed chicks produced louder cluster 1 calls, but quieter cluster 0 calls in later time bins. This bidirectional pattern may reflect altered energy regulation in VPA-exposed chicks, with reduced intensity in softer calls and increased intensity in contact calls. These findings partially align with Nishigori et al. \cite{nishigori2013impaired}, where VPA-exposed chicks showed an overall call amplitude reduction compared to controls. However, methodological differences make direct comparisons difficult. Nishigori et al. \cite{nishigori2013impaired} did not differentiate call types and analysed only call decibels. Further, they recorded chicks socially isolated after imprinting with conspecifics (post-hatching day P2--P4), whereas we tested newly hatched chicks (P0). Their overall intensity reduction may therefore reflect specific call type changes, consistent with our cluster 0 energy reduction (Figure~\ref{fig:energy_features}), which includes ``soft calls'' (see Table~\ref{tab:literature_computational_comparison}).

In the frequency domain, group differences in fundamental frequency variability emerged. In VPA-exposed chicks, cluster 1 calls showed lower F0 Standard Deviation, F0 Slope, and F0 1st Order Difference, indicating reduced pitch variability and dynamic modulation compared to controls. The F0 Kurtosis was higher in VPA-exposed chicks for both clusters, reflecting less variable pitch distribution and flatter calls, with effects most pronounced in cluster 1. 

Finally, Spectral Centroid Standard Deviation was reduced in VPA-exposed chicks in both clusters, indicating decreased spectral variability. These alterations align with rodent models, where VPA exposure led to reduced call complexity and shifts toward simpler call types with altered peak frequency \cite{gzielo2020valproic, moldrich2013inhibition, nicolini2018valproic} and indicate call-type-specific effects.
 % Findings' Scientific Implications for biological side
These findings indicate that the impact of prenatal VPA exposure is not uniform across the vocal repertoire, and may reflect alterations in social communication paralleling those observed in social behaviour \cite{sgado2018embryonic, lorenzi2019embryonic, adiletta2021spontaneous}.

Since cluster 1 represents calls with greater acoustic variability and F0 modulation dynamics, further investigation is needed to determine whether selective alterations originate from peripheral vocal production differences, motor impairments, or socio-cognitive and motivational deficits. Our experimental dataset was recorded during a social preference test with socially relevant stimuli, potentially eliciting specific vocalisation patterns. Interestingly, the average stimulus approach time (170--190 seconds) \cite{lorenzi2019embryonic} coincides with increased call duration and intensity for cluster 1 calls (Figures \ref{fig:temporal_features} and \ref{fig:energy_features}), suggesting connections between social and motor responses and vocalisations to be explored in future work.

\section{Conclusion}

Our computational framework addresses two major bottlenecks in chicks' vocal research: the labour-intensive nature of manual annotation and the subjective biases in human-defined classification systems.
Being automated and reproducible, our pipeline can efficiently process large datasets, making it suitable for longitudinal studies and high-throughput applications in neuroscientific research and precision livestock farming. 
Furthermore, the framework can be applied to secondary data, uncovering previously unexamined acoustic dimensions in existing datasets not originally collected for acoustic analysis.

% Forward-looking statement about the field as a whole.
Extended to other species and experimental contexts, this methodology could facilitate cross-species comparisons of vocal communication and its disruption. As the field is moving toward more interdisciplinary approaches \cite{sainburg2021toward, robinson2024naturelm}, integrating acoustic analysis with neural substrate and behavioural tracking can help to understand the vocal and communicative dimensions of developmental disorders.

% Limitations of the Platform
Some limitations warrant consideration. Six-minute recordings on the first day of hatching during a social preference test may have captured specific vocalisation patterns, not reflecting the entire variety of vocal behaviour in these social animals. 
Future studies should investigate whether alterations persist throughout development and in different contexts, and further validate the framework on larger and more diverse datasets, spanning different ages, acoustic environments, and background-noise conditions.
Although our feature set covered the temporal, frequency, and energy domains, higher-order features such as joint time-frequency scattering coefficients \cite{wang2022joint} could capture nonlinear acoustic interactions and reveal further vocal differences. From a neuroscientific perspective, these results highlight how fine-grained vocal analysis can detect subtle phenotypes in animal models of neurodevelopmental disorders and inform the investigation of neural substrates underlying communication alterations. As candidate vocal markers of neurodevelopmental disruption, future studies should focus on call duration, energy, and the variability of pitch and spectral shape, which showed significant differences between VPA-exposed and control chicks in our data. Avian models provide a well-suited system for linking acoustic phenotypes to neural substrates and circuit-level mechanisms \cite{csillag2022avian}. These results reinforce the value of domestic chicks as a model for studying innate vocal behaviour and highlight the potential of computational bioacoustics for understanding neurodevelopmental conditions \cite{matsushima2024domestic, adiletta2021spontaneous, zachar2019valproate}.

\section*{Data availability}
The dataset is publicly available via Zenodo (\url{https://doi.org/10.5281/zenodo.19468051}). The code is available on GitHub (\url{https://github.com/antorr91/VPA_vocalisations_project}) and archived on Zenodo (\url{https://doi.org/10.5281/zenodo.23641590}).

\section*{Authors' contributions}

AMCT: Conceptualisation, Data curation, Formal analysis, Methodology, Software, Visualisation, Writing. 
IN: Software, Writing (review). 
PS: Conceptualisation, Data curation, Resources, Writing (review). 
EV: Conceptualisation, Methodology, Funding acquisition, Supervision, Writing (review). 
EB: Conceptualisation, Methodology, Funding acquisition, Project administration, Supervision, Software, Writing (review).

\section*{Funding}

IN has been supported by a BBSRC grant (BB/S003223/1); PS has been supported by the Italian Ministry of University and Research (PS, Grant number 2022LJZRBY); EV has been supported by a Royal Society Leverhulme Trust Senior Research Fellowship (SRF\textbackslash R1\textbackslash 210001155), a Leverhulme Trust grant (RPG-2020-287) and a BBSRC grant (BB/S003223/1); EB has been supported by a RAEng/Leverhulme Trust Research Fellowship (LTRF2223-19-106).

\section*{Ethics statement}

Experimental procedures adhered to national regulations on animal research. For the experimental dataset, procedures were approved by the University of Trento ethical committee and authorised by the Italian Ministry of Health (permit no. 1060/2015-PR), in accordance with EU Directive 2010/63/EU. For the Development dataset, procedures were approved by the Queen Mary University of London ethics committee (AWERB) and Home Office (PP5180959).

\section*{Acknowledgements}

We thank Michael Emmerson, Laura Freeland, and Shuge Wang for their assistance with the annotation of the Development dataset.

\bibliographystyle{RS} 

% \bibliography{bibliography_interface}

\newpage

\appendix

\setcounter{figure}{0}
\setcounter{table}{0}
\renewcommand{\thesection}{S\arabic{section}}
\renewcommand{\thefigure}{S\arabic{figure}}
\renewcommand{\thetable}{S\arabic{table}}

\section*{Supplementary material}

\small 

\section*{Extended methods}
\label{sec:extended_methods}
This section provides detailed methodological procedures that complement the main Methods section.

\section{Onset and Offset Detection}
\label{sec:extended_method_onset_offset}

We used the best onset and offset detection algorithms identified in a previous study conducted on the development dataset \cite{torrisi2024exploratory}.

For onset detection, the High Frequency Content (HFC) algorithm \cite{muller2015fundamentals} was selected based on its superior performance compared to other algorithms tested, including phase and amplitude-based methods (Thresholded Phase Deviation, Normalised Weighted Phase Deviation, and Rectified Complex Domain \cite{muller2015fundamentals}) and energy-based methods like Superflux \cite{bock2013maximum}. The HFC algorithm detects frequency changes within the signal, making it particularly suited for identifying the sharp spectral transitions characteristic of chick vocalisation onsets.

For offset detection, we implemented three methods operating within a predefined time window based on the maximum expected duration of chick calls: (i) local minimum detection of the energy signal, (ii) first-order differencing of the energy signal followed by local minimum detection, and (iii) second-order differencing combined with local minimum detection. The first-order differencing method was selected based on its superior performance compared to the other approaches tested. Performance evaluation on the development dataset and benchmarking results on the experimental dataset are presented in Section~\ref{sec:s1_onset_offset_benchmarking}.

\subsection{Onset and offset detection performance}
\label{sec:s1_onset_offset}

\subsubsection{Evaluation metrics}

To evaluate onset and offset detection performance, predicted events were compared to manually annotated ground-truth labels using event-based metrics \cite{mesaros2016metrics}. A prediction was considered a True Positive (TP) if it fell within a defined tolerance window relative to the ground-truth.
For the onset detection task, a fixed tolerance window of 100 ms (±50 ms) was used, based on preliminary qualitative analysis that confirmed this threshold as optimal for chick calls. 
For the offset detection task, an adaptive tolerance window was used. While a fixed 100 ms window was applied by default, for longer events the tolerance was set to half the total event duration (e.g., 150 ms for a 300 ms call), to account for the greater temporal variability typically associated with offset annotations in vocalisations of variable length. Predicted offsets were estimated using the corresponding predicted onset as an anchor and evaluated call-wise against the corresponding ground truth offsets.
Undetected vocalisations were counted as False Negatives (FN), while extra or duplicated detections were considered False Positives (FP).
Evaluation metrics were computed using the \texttt{onset.evaluate} function from the \texttt{mir\_eval} Python library \cite{raffel2014mir_eval}:
\begin{enumerate}
 \item \textbf{Precision}: Represents the ratio of True Positives against all instances estimated as positive, including False Positives
 \[
 \text{Precision} = \frac{\text{True Positives}}{\text{True Positives + False Positives}}
 \]
 \item \textbf{Recall (Sensitivity)}: Ratio of True Positives to all actual positives (True Positives + False Negatives). 
 \[
 \text{Recall} = \frac{\text{True Positives}}{\text{True Positives + False Negatives}}
 \] 
 \item \textbf{F1-measure}: Combines Precision and Recall, producing a unified measure (the harmonic mean). The F1-measure provides insight into both the model's capacity to make accurate positive predictions and its capacity to capture all real positive instances.
 \[
 \text{F1} = \frac{2 \cdot \text{Precision} \cdot \text{Recall}}{\text{Precision} + \text{Recall}}
 \]
\end{enumerate}

These metrics were computed individually for each audio file and aggregated using weighted averages to handle the imbalanced distribution of calls within each audio file:
\[
\text{Metric}_{\text{weighted}} = \sum_{i=1}^{N} w_i \cdot \text{Metric}_i
\]

\subsubsection{Results detection performance across datasets}
\label{sec:s1_onset_offset_benchmarking}
To ensure the robustness and generalisability of the automatic detection system, onset and offset detection were evaluated on both the development dataset and on the experimental dataset for benchmarking. In Table~\ref {tab:onset_offset_performance}, detection performances are presented across both datasets.

\begin{table}[htbp]
\centering
\renewcommand{\arraystretch}{1.6}
\resizebox{1\linewidth}{!}{ 
\begin{tabular}{llccc}
\hline
\textbf{Dataset} & \textbf{Detection Type} & 
\shortstack{\rule{0pt}{2.6ex}\textbf{Weighted}\\\textbf{F1-measure}} & 
\shortstack{\rule{0pt}{2.6ex}\textbf{Weighted}\\\textbf{Precision}} & 
\shortstack{\rule{0pt}{2.6ex}\textbf{Weighted}\\\textbf{Recall}} \\
\hline
Development Dataset (31 naïve chicks) & Onset (HFC) & 0.853& 0.908& 0.841\\
Development Dataset (31 naïve chicks) & Offset (1st-order difference) & 0.724& 0.724& 0.724\\
\hline
Experimental Dataset (46 chicks: 27 VPA / 19 control) & Onset (HFC) & 0.933& 0.910& 0.961\\
Experimental Dataset (46 chicks: 27 VPA / 19 control) & Offset (1st-order difference) & 0.885 & 0.885 & 0.885 \\
\hline
\end{tabular}
}
\caption{Performance metrics for automatic onset and offset detection across datasets. HFC = High Frequency Content algorithm; 1st-order difference = First order difference of energy + local minimum detection.}
\label{tab:onset_offset_performance}
\end{table}

The results confirmed the effectiveness of the automatic detection methods on the complete experimental dataset. Onset detection performed consistently well across datasets, with improved performance in the experimental Dataset (F1 = 0.933) compared to the development Dataset (F1 = 0.853). Similarly, offset detection showed substantial improvement in the experimental Dataset (F1 = 0.885) compared to the development Dataset (F1 = 0.724), likely reflecting better signal quality for the audio recorded for the experimental dataset. These results validated the use of automated detection in this study. The high precision and recall values demonstrate the reliability of our method for large-scale signal analysis.

\section{Features extracted for the chicks' calls analysis}
\label{sec:extended_method_feature_extraction}

Before feature extraction, a bandpass filter (BPF) was applied to each recording to reduce background noise and focus on the frequency bands relevant to chick vocalisations, specifically the fundamental frequency along with its first and second harmonics. The frequency range of the filter was set to 2000--12,600 Hz for the development dataset and to 2000--15,000 Hz for the experimental dataset, as preliminary visual inspection indicated that the chicks' formants reached slightly higher frequencies in the experimental recordings; therefore, the upper bound was set to preserve harmonic content.

For each segmented call, we extracted different components of the signal. The Fundamental Frequency (F0), indicating the average number of oscillations per second (in Hz), was estimated using the PYIN algorithm \cite{mauch2014pyin}. From this estimate, frequency bins corresponding to the first and second harmonics (F1 and F2) were derived \cite{muller2015fundamentals}, and their magnitudes computed via the Fast Fourier Transform (FFT), with normalisation relative to the sampling frequency. We then computed the Root Mean Square (RMS) \cite{Panagiotakis}, which measures the signal energy and provides an estimate of its overall intensity, and the Spectral Centroid, which represents the weighted average of frequencies in the signal spectrum. For the Spectral Centroid \cite{muller2015fundamentals}, we focused on the typical frequency band of chick vocalisations, between 2000 Hz and 12,600 Hz. The signal envelope \cite{virtanen2018computational} was derived by segmenting the waveform of each call and applying the Hilbert transform to compute the analytic signal, from which the absolute amplitude was extracted. 

From these signal-level features, we characterised each call by calculating 20 one-dimensional statistical descriptors. These are grouped into: temporal domain features (Call Duration, Attack Time, Envelope Slope), frequency domain features (F0 Mean, F0 Standard Deviation, F0 Kurtosis, F0 Skewness, F0 Bandwidth, F0 Slope, F0 1st Order Difference Mean, F0 Mag Mean, F1 Mag Mean, F2 Mag Mean, F1/F0 Ratio Mean, F2/F0 Ratio Mean, Spectral Centroid Mean and Spectral Centroid Standard Deviation), and energy domain features (RMS Mean, RMS Standard Deviation and Attack Magnitude). Each descriptor is a within-call statistic, computed from the frame-level values of the corresponding signal feature. 

Detailed feature descriptions and equations to derive them are provided in Table~\ref{tab:features_and_equations}. The implementation code used to derive these features is publicly accessible at the following link\footnote{\url{https://github.com/antorr91/Chicks_exploratory_study/blob/main/feature_extraction_functions.py}}, supporting transparency and reproducibility of the signal processing pipeline. 
During feature extraction, 23 detections in the experimental dataset (11 VPA, 12 control) with values outside the range of chick calls were excluded as noise, as confirmed by visual inspection of their spectrograms. 
All extracted features were z-score normalised prior to clustering, ensuring they were on a comparable scale for these analyses.

\begin{table}[htbp]
 \centering
 \footnotesize
 \renewcommand{\arraystretch}{1.8}
 \begin{tabular}{p{2.6cm}p{6.5cm}p{3.2cm}}
 \toprule
 \textbf{Feature} & \textbf{Definition} & \textbf{Equation} \\
 \midrule
 Call Duration (T) & Time difference between the offset and onset of a call. & $T_{\text{end}} - T_{\text{start}}$ \\
 
 F0 Mean (F) & Average fundamental frequency within the call. & $\frac{1}{N} \sum_{i=1}^N F0_i$ \\

 F0 Std Dev (F) & Variability of fundamental frequency values within the call.& $\sqrt{\frac{1}{N} \sum_{i=1}^N (F0_i - \overline{F0})^2}$ \\
 
 F0 Skewness (F) & Symmetry of the F0 distribution (positive = right-skewed).& $\frac{\frac{1}{N} \sum_{i=1}^N (F0_i - \overline{F0})^3}{\left(\sqrt{\frac{1}{N} \sum_{i=1}^N (F0_i - \overline{F0})^2}\right)^3}$ \\
 
 F0 Kurtosis (F) & Peakedness or flatness of the F0 distribution (positive value = heavier tails, irregular modulation; negative value = lighter tails, smoother modulation). & $\frac{\frac{1}{N} \sum_{i=1}^N (F0_i - \overline{F0})^4}{\left(\sqrt{\frac{1}{N} \sum_{i=1}^N (F0_i - \overline{F0})^2}\right)^4} - 3$ \\
 
 F0 Bandwidth (F) & Range or span of F0 values within a call. & $\max(F0) - \min(F0)$ \\
 
 Mean of F0's First-Order Difference (F) & Average of differences between consecutive F0 values. & $\frac{1}{N-1} \sum_{i=2}^N (F0_i - F0_{i-1})$ \\

 F0 Slope (F) & Rate of frequency change from onset to peak. & $\frac{F0_{\text{peak}} - F0_{\text{onset}}}{t_{\text{peak}} - t_{\text{onset}}}$ \\
 
 F0 Magnitude Mean (F) & Average intensity of the fundamental frequency within the call. & $\frac{1}{N} \sum_{i=1}^N |X(F0_i)|$ \\
 
 F1 Magnitude Mean (F) & Average intensity of the first harmonic within the call. & $\frac{1}{N} \sum_{i=1}^N |X(F1_i)|$ \\
 
 F2 Magnitude Mean (F) & Average intensity of the second harmonic within the call. & $\frac{1}{N} \sum_{i=1}^N |X(F2_i)|$ \\

 Ratio F1/F0 Magnitude Mean (F) & Mean of the ratio between the F1 and F0 magnitudes. & $\frac{1}{N} \sum_{i=1}^N \frac{|X(F1_i)|}{|X(F0_i)|}$ \\

 Ratio F2/F0 Magnitude Mean (F) & Mean of the ratio between the F2 and F0 magnitudes. & $\frac{1}{N} \sum_{i=1}^N \frac{|X(F2_i)|}{|X(F0_i)|}$ \\

 Spectral Centroid Mean (F) & Average spectral centroid within the call. & $\frac{1}{N} \sum_{i=1}^N \frac{\sum_{k=1}^K f_k \cdot |X(f_k,i)|}{\sum_{k=1}^K |X(f_k,i)|}$ \\

 Spectral Centroid Std Dev (F) & Variability of spectral centroid values within the call. & $\sqrt{\frac{1}{N} \sum_{i=1}^N \left(SC_i - \overline{SC}\right)^2}$ \\

 RMS Mean (E)& Average RMS value indicating overall energy within the call. & $\frac{1}{N} \sum_{i=1}^N RMS_i$ \\
 
 RMS Std Dev (E)& Variability of RMS values within the call. & $\sqrt{\frac{1}{N} \sum_{i=1}^N (RMS_i - \overline{RMS})^2}$ \\
 
 Attack Magnitude of the Envelope (E) & Intensity of the attack of the call. & $|x(t_{\text{peak}})| - |x(t_{\text{onset}})|$ \\
 
 Attack Time of the Envelope (T) & Duration from onset to peak amplitude. & $t_{\text{peak}} - t_{\text{onset}}$ \\
 
 Slope of the Envelope (T) & Rate of change of amplitude during attack. & $\frac{|x(t_{\text{peak}})| - |x(t_{\text{onset}})|}{t_{\text{peak}} - t_{\text{onset}}}$ \\
 \bottomrule
 \end{tabular}
\caption{Extracted features for the exploratory study. (T) = Time-domain, (E) = Energy domain, (F) = Frequency-domain. Each feature is computed within a single call from its frame-level values, where $N$ is the number of frames in the call, $i$ indexes the frames, and $|X(f)|$ is the FFT magnitude at frequency $f$. Features such as Call Duration, F0 Bandwidth, F0 Slope, Attack Magnitude, Attack Time and Envelope Slope are not derived from frame-level summaries.}

 \label{tab:features_and_equations}
\end{table}

\subsection{Acoustic and biological interpretation of features.}
\label{features_interpretation}

The 20 features summarised in Table~\ref{tab:features_and_equations} were selected by building on previous literature on the characterisation of the acoustic features most relevant for describing and distinguishing different vocalisations (see the work on Predefined Acoustic Features by Elie and Theunissen \cite{elie2016vocal}). These acoustic features are also relevant for conveying important communicative and affective value \cite{briefer2012vocal, manteuffel2004vocalization}. These features can therefore reveal not only information relevant to chick welfare \cite{collins2024sound}, but also key insights into the biological differences observed in neurodevelopmental models, in which alterations of social communication have already been reported in other species \cite{mimura2026altered, nicolini2018valproic, tewelde2025prenatal}. In chicks, before the current study, a lower overall amplitude in vocalisation among VPA-exposed chicks had been documented \cite{nishigori2013impaired}. Below, we summarise the interpretation of the features we analysed across the temporal, frequency, and energy domains. 

We quantified the temporal domain through the feature Call duration. This feature is consistently linked to affective state: it increases with arousal across species \cite{briefer2020coding, elie2016vocal}, and the same relationship has been documented directly in domestic chicks, where distress calls become longer as arousal increases \cite{collins2024sound}. Attack Time and Envelope Slope capture the dynamics of vocal onset, that is, the speed and slope with which the signal reaches its peak. These features reflect the tension in the respiratory tract \cite{briefer2020coding}, indicating the temporal and motor control of vocal production. Lower values in these aspects could either signal lower arousal states or lower vocal control. Consistent with this, prenatal VPA exposure in rodent models reduces call duration and shifts the repertoire towards shorter, simpler calls \cite{gzielo2020valproic}.

In the energy domain, the RMS Mean and RMS standard deviation measure the overall intensity of the call and the within-call variation in vocal effort, while Attack Magnitude captures the intensity of the call's transient from the onset to the peak. Louder calls are linked to higher arousal levels across mammals \cite{briefer2020coding, elie2016vocal}. This has been confirmed in chicks, where higher-arousal calls are louder and energetic \cite{collins2024sound}. Through our analysis, we measured both the average energy and its variability, which allows for a nuanced characterisation of energy modulation specific to different call types.

In the frequency domain, we characterised both the level of pitch, through F0 Mean, and how pitch varies and is modulated across the call, through F0 Standard Deviation, F0 Bandwidth, F0 Slope, F0 1st Order Difference, F0 Kurtosis and F0 Skewness. These last capture the shape of the pitch distribution and the rate of pitch change over the call, and thus measure prosodic complexity. The pitch overall correlates with arousal: across species, including domestic chicks, higher-arousal states are linked to higher F0 Mean and greater F0 range and variability \cite{briefer2020coding, elie2016vocal, collins2024sound}. Including pitch variability descriptors in our analysis helped us to describe a call's modulation, and we found that lower values in these features corresponded to flatter, less modulated and more stereotyped calls in VPA-exposed chicks.

Finally, within the frequency domain, we quantified spectral shape and formants through Spectral Centroid Mean and Spectral Centroid Standard Deviation, the F0, F1 and F2 magnitudes, and the F1/F0 and F2/F0 ratios. 

The Spectral Centroid is the amplitude-weighted mean of the frequency spectrum, i.e., its centre of mass. It reflects the call's predominant frequency content, and its perceptual, psychoacoustic correlate is the timbral brightness \cite{marozeau2007effect}, with higher values indicating a brighter sound and lower values a darker one. The F0, F1 and F2 magnitudes and their ratios describe the level of the energy between the fundamental frequency and its harmonics. The shape of the spectral envelope is one of the most important features in discriminating call types across species \cite{elie2016vocal}, and it also relates to affective valence: a noisy spectral quality has been associated with negative states in domestic fowl, and in chicks it is captured by spectral entropy and the harmonics-to-noise ratio \cite{collins2024sound}.

Emotional valence is less consistently expressed than arousal and has been observed to be species-specific \cite{briefer2020coding, laurijs2021vocalisations}. Cross-species findings indicate that positive states are associated with shorter Call Duration, with larger amplitude modulations of the RMS features in some species, with lower F0 Mean and reduced F0 variability, and with changes in spectral shape, where a more tonal sound tends to accompany positive states and darker and noisier with negative states in domestic fowl \cite{briefer2020coding, laurijs2021vocalisations}. Finally, it is important to note that sound production varies within a species according to individual variables such as body size and vocal tract, regardless of affective state \cite{dunn2026constraint}, so these interpretations are indicative rather than definitive, and a one-to-one correspondence between features, affective valence and other individual variables such as body size and sex would require a controlled study that lies beyond the scope of the present work.

\section{Clustering analysis}
\label{sec:clustering_supplementary}

For the analysis of chicks’ vocal repertoire, we evaluated hard and soft clustering methods to capture potentially different underlying structures. Hard clustering techniques, where each instance belongs to only one cluster, are well-suited for identifying discrete categories \cite{fischer2017structural}. We tested three hard clustering techniques: K-means \cite{rokach2023machine}, Hierarchical Agglomerative Clustering (HAC)\cite{rokach2023machine}, and Density-Based Spatial Clustering of Applications with Noise (DBSCAN) \cite{rokach2023machine}. In contrast, soft clustering techniques allow each instance to belong to multiple clusters, making them more appropriate when the data exhibit a continuous distribution\cite{zadeh2008there}. We tested two soft clustering techniques: Fuzzy C-Means \cite{rokach2023machine} and Gaussian Mixture Models (GMMs) \cite{rokach2023machine}. Both families were tested in an initial comparison, so that the choice of method could be based on the observed performance rather than on assumptions about the underlying structure of the vocal repertoire.

To identify the optimal clustering structure for the development dataset, we performed a grid search, testing solutions from 1 to 10 clusters for each algorithm. Three metrics guided the selection of the number of clusters: (i) Silhouette Score \cite{rokach2023machine}: quantifies how similar each point is to its own cluster compared to others; (ii) Within Cluster Sum of Squares (WCSS) \cite{rokach2023machine} used with the Elbow method to identify the point beyond which adding more clusters does not significantly improve compactness; (iii) Calinski-Harabasz Index (CHI, or Variance Ratio Criterion, VRC) \cite{rokach2023machine}: evaluates the ratio of dispersion between and within clusters. For soft clustering, we used additional metrics: Fuzzy Partition Coefficient (FPC)\cite{rokach2023machine}, which evaluates the quality of fuzzy separation (for Fuzzy C-Means); Akaike Information Criterion (AIC) \cite{rokach2023machine} and Bayesian Information Criterion (BIC) \cite{rokach2023machine} that balance model fit and complexity (for GMM). The single-cluster solution was evaluated through WCSS and, for Gaussian Mixture Models, through AIC and BIC, as the Silhouette Score and CHI require at least two clusters (see Supplementary Section~\ref{sec:clustering_results_development_dataset}).
We considered these metrics jointly, as each captures a different aspect of cluster structure \cite{goffinet2021low}. In graded vocal repertoires, silhouette values can be low even in the presence of meaningful structure \cite{fischer2017structural, sainburg2020finding}, supporting the use of multiple metrics.

After identifying the best cluster division, we inspected calls at varying distances from the cluster centroids to assess within-cluster homogeneity and identify potential outliers, using a percentile-based sampling method: calls within each cluster were ranked by Euclidean distance from centroids, which represent the mean acoustic profile of each cluster. Given its higher internal consistency and lower number of outliers, we report the results of HAC. This may reflect its lower sensitivity to non-uniform cluster sizes and densities compared with centroid-based methods, which tend to favour partitions of comparable size, a relevant consideration given the marked imbalance observed between the two vocal clusters.

For HAC, we employed Ward linkage, which minimises the increase in within-cluster variance when merging clusters and is suited for a multi-dimensional acoustic feature space. 
The same grid search approach was then applied to the experimental dataset using HAC and evaluated through the Silhouette Score, CHI, and WCSS.
To visualise this hierarchical clustering structure for both datasets, we generated dendrograms, where branch height represents the Ward distance at which clusters are merged.

\subsection{Clustering analysis results in the development dataset}
\label{sec:clustering_results_development_dataset}

This section summarises the comparative evaluation of clustering methods applied to the vocalisations of naïve chicks, recorded 12 hours after hatching. Although grid search optimisation was performed for cluster numbers ranging from \( K = 1 \) to \( K = 10 \), only the most informative solutions (\( K \leq 5 \)) are reported here, as higher values consistently yielded diminishing returns across all metrics. 

Clustering performance was assessed using the internal validity metrics described in the Methods section. The results for each algorithm and cluster number are summarised in Table ~\ref{tab:summary_metrics}.

\begin{table}[h]
\centering
 \scriptsize
\footnotesize
\renewcommand{\arraystretch}{1.5}
\begin{tabular}{p{1.18cm}ccccl}
\toprule
\textbf{Cluster} & \textbf{Method} & \textbf{Silhouette Score} & \textbf{CHI} & \textbf{WCSS} & \textbf{Soft clustering metrics}\\
\midrule
1 & K-means & -- & -- & 118480.0000 & - \\
1 & Fuzzy C-means & -- & -- & 118480.0000 & FPC: 1.000000\\
1 & HAC & -- & -- & 118480.0000 & - \\
1 & GMM & -- & -- & 118480.0000 & BIC: 224507.30, AIC: 222969.34\\
1 & DBSCAN & -- & -- & 105558.2274 & - \\
\midrule
2 & K-means & \textbf{0.293544} & \textbf{2075.165721} & 87736.037814 & - \\
2 & Fuzzy C-means & \textbf{0.292199} & \textbf{2064.470661} & 57202.438842 & \textbf{FPC: 0.618044}\\
2 & HAC & \textbf{0.262055} & \textbf{1820.717257} & \textbf{90619.163365} & - \\
2 & GMM & 0.163535 & 665.411260 & 106512.447792 & BIC: 85477.93, AIC: 82395.33\\
2 & DBSCAN & \textbf{0.612587} & 32.850100 & 103535.413600 & - \\
\midrule
3 & K-means & 0.205640 & 1444.535753 & \textbf{79627.221719} & - \\
3 & Fuzzy C-means & 0.203000 & 1425.761100 & \textbf{37624.026316} & FPC: 0.423804\\
3 & HAC & 0.173450 & 1214.278693 & 84018.834471 & - \\
3 & GMM & \textbf{0.189538} & 897.085339 & 90928.248410 & \textbf{BIC: 44965.50}, \textbf{AIC: 40338.25}\\
3 & DBSCAN & 0.541936 & \textbf{36.791861} & 102265.344510 & - \\
\midrule
4 & K-means & 0.218570 & 1242.981469 & 72692.093594 & - \\
4 & Fuzzy C-means & 0.085100 & 931.119700 & 28213.933835 & FPC: 0.318023\\
4 & HAC & 0.175407 & 989.556403 & 78909.630165 & - \\
4 & GMM & 0.175460 & \textbf{948.214865} & \textbf{80026.578493} & BIC: 18210.01, AIC: 12038.12\\
4 & DBSCAN & 0.480605 & 16.904633 & 101784.735236 & - \\
\midrule
5 & K-means & 0.185797 & 1087.206512 & 68299.126302 & - \\
5 & Fuzzy C-means & 0.019200 & 647.308900 & 22638.646939 & FPC: 0.253115\\
5 & HAC & 0.132163 & 893.086602 & 73886.579398 & - \\
5 & GMM & 0.087556 & 740.931014 & 78949.551308 & BIC: 4527.62, AIC: -3188.90\\
5 & DBSCAN & 0.366437 & 12.984974 & 97517.557780 & - \\
\bottomrule
\end{tabular}
\caption{Clustering performance metrics for \(K=1\) to \(K=5\). FPC = Fuzzy Partition Coefficient; AIC = Akaike Information Criterion; BIC = Bayesian Information Criterion; CHI = Calinski-Harabasz Index; WCSS = Within-Cluster Sum of Squares. Bold values indicate optimal scores per metric. The optimal number of clusters for WCSS, AIC, and BIC was identified using the Elbow method.}
\label{tab:summary_metrics}
\end{table}

The clustering results indicate that the optimal number of clusters varies across clustering methods and evaluation metrics.

For \(K=1\), the Silhouette Score and CHI cannot be computed, as both metrics measure the separation between clusters. The Silhouette Score compares, for each sample, the mean intra-cluster distance with the mean distance to the nearest other cluster, and the CHI is the ratio of the between-cluster dispersion to the within-cluster dispersion \cite{rokach2023machine}. WCSS can instead be computed; for a single cluster, the centroid is the mean of all calls, so WCSS equals the total sum of squares of the data. For this reason, WCSS is identical across methods, except for DBSCAN, where it is computed excluding noise points.  For GMM, AIC and BIC could also be computed, whereas for Fuzzy C-means (FCM) the FPC equals 1 by definition, as all calls fully belong to the single cluster.

For \(K=2\), K-means, FCM, and HAC show strong internal consistency, with the highest Silhouette and CHI values across all the numbers of clusters tested (Silhouette: 0.2621–0.2935, CHI: 1820.717–2075.166). HAC also shows the optimal partition with WCSS, as the two-cluster solution produces the largest reduction among all the tested solutions. For FCM, this cluster division is also supported by the peak value of FPC (0.6180). 
Results from the other two methods partially diverge. GMM shows lower Silhouette and CHI (0.1635 and 665.4113). DBSCAN reaches the highest Silhouette (0.6126), but a lower CHI than the other methods (32.8501), indicating that it captures local densities rather than the global structure of chicks' vocal repertoire.

For \(K=3\), GMM improves, reaching its highest Silhouette (0.1895) and good CHI (897.0853). At this partition, GMM also reaches its best balance between cluster quality and model complexity (AIC: 40338.25, BIC: 44965.50). In contrast, Silhouette and CHI progressively decrease for HAC and the centroid-based methods (K-means, FCM) (Silhouette: 0.1734--0.2056, CHI: 1214.28--1444.536), while K-means and FCM show the optimal partition with WCSS.

For \(K=4\), GMM shows the optimal partition with WCSS and reaches its highest CHI (948.21). However, the Silhouette score does not improve at this partition for any method, indicating that the additional cluster does not enhance cluster separation.

The difference in the Silhouette and CHI values between the distance-based methods (K-means, FCM, HAC) and the other two lies in the continuous structure of the data (Figure~\ref{fig:clustering_umap_dev_dataset}); indeed, in a graded distribution, likelihood- and density-based criteria can favour additional components even without a genuine separation between groups.
Overall, these findings support that the selection of a two-cluster solution provides the best balance between separation and complexity.

\subsection{Clustering analysis results in the experimental dataset}
\label{sec:clustering_results_experimental_dataset}

We applied Hierarchical Agglomerative Clustering (HAC) independently to the VPA and control groups in the experimental dataset. For the VPA-treated group, the Silhouette Score peaked at \( K = 2 \) (0.425), indicating that vocalisations were best distinguished at a broader categorical level. The CHI peaked at K = 2 (1724.68) and declined thereafter, further supporting the presence of two primary vocalisation categories. However, the Elbow method applied to WCSS identified \( K = 4 \) (WCSS = 219535.80) as the optimal number of clusters, suggesting a possible substructure in the data. Despite this, the relatively low Silhouette Score at \( K = 4 \) (0.125) suggested poor cluster separation. This is consistent with a more continuous organisation of vocal features rather than sharply defined categories.

\begin{figure}[h]
 \centering
 \includegraphics[width=\linewidth]{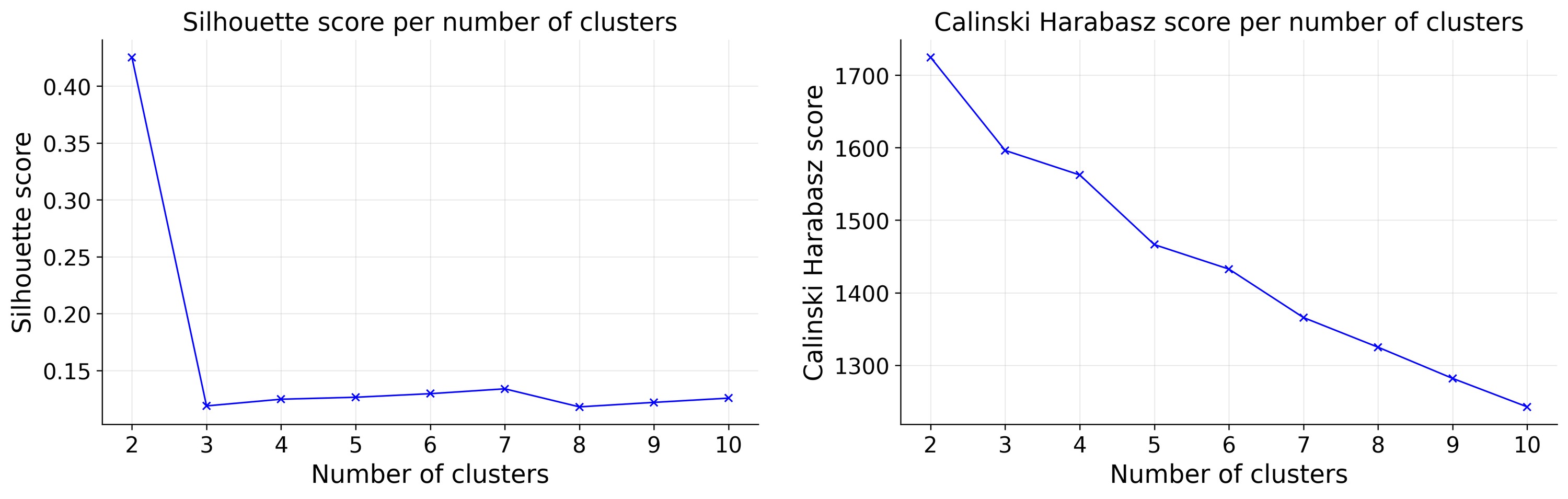}
 \caption{Clustering performance metrics for the VPA group across cluster solutions (\(K=2\) to \(K=10\)). Silhouette Score and CHI peak at \(K=2\).}
 \label{fig:VPA_silhouette_chi}
\end{figure}

\begin{figure}[h]
 \centering
 \includegraphics[width=0.65\linewidth]{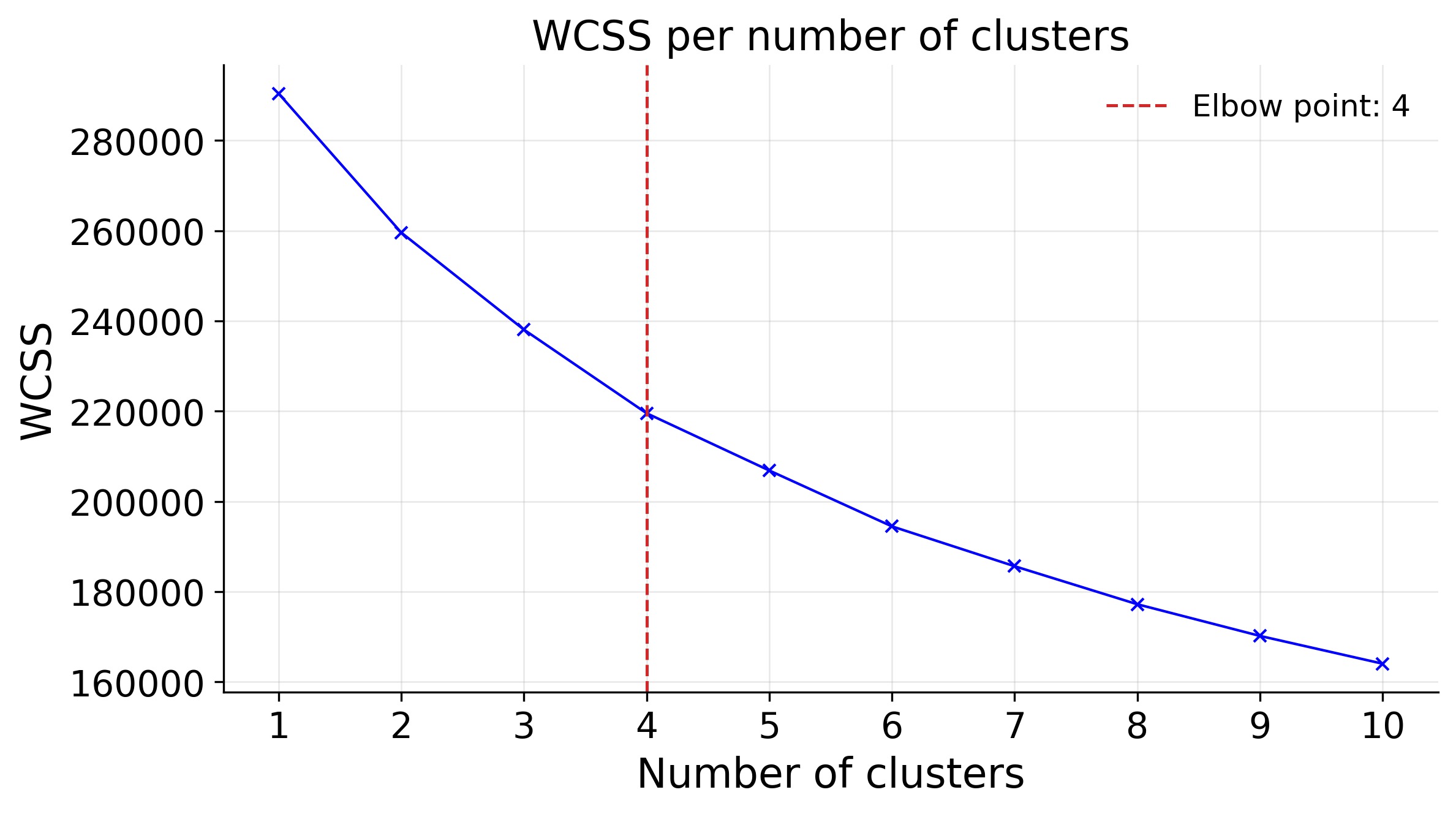}
 \caption{Within-Cluster Sum of Squares (WCSS) for the VPA group. The Elbow method suggests an inflexion at \(K=4\).}
 \label{fig:VPA_wcss}
\end{figure}

Similarly, for control chicks, the highest Silhouette Score was observed at \( K = 2 \) (0.393), while \( K = 3 \) (0.347) also provided a reasonable balance between separation and model complexity. The CHI index was highest at \( K = 2 \) (1820.28), indicating strong structural differentiation at this level. Although the Elbow method indicated that \( K = 3\)  (WCSS = 137809.06) was optimal based on WCSS trends, the Silhouette Score at \( K = 3 \) (0.347) remained lower than at \( K = 2 \).

\begin{figure}[h]
 \centering
 \includegraphics[width=\linewidth]{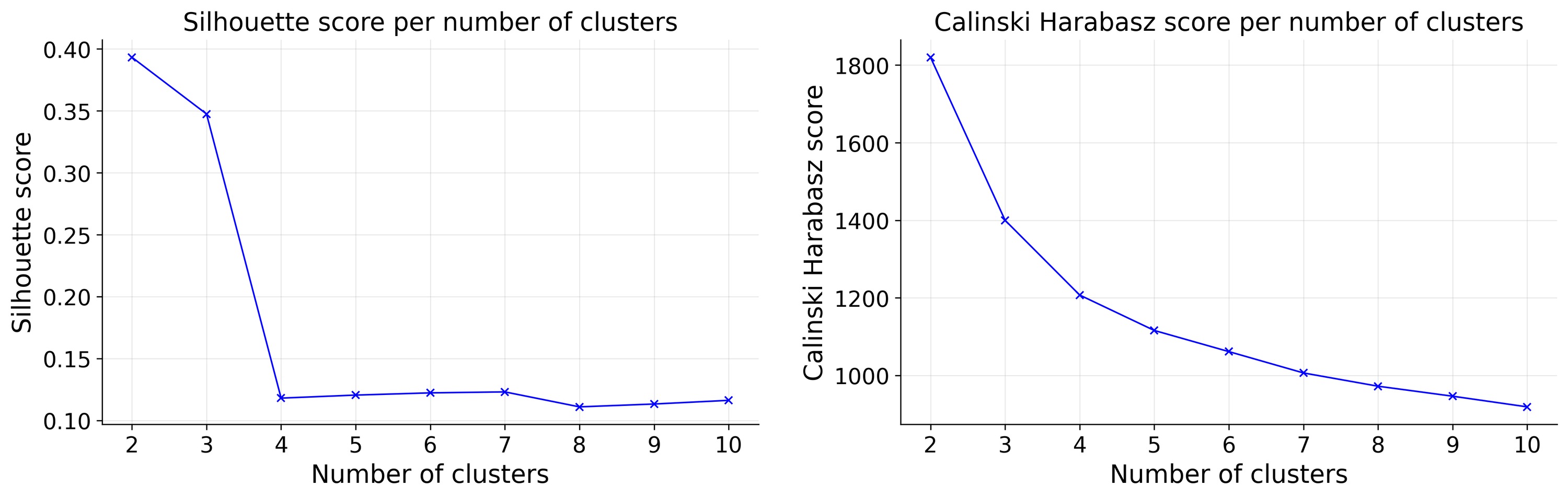}
 \caption{Clustering performance metrics for the control group across cluster solutions (\(K=2\) to \(K=10\)). Both Silhouette Score and CHI are highest at \(K=2\).}
 \label{fig:control_silhouette_chi}
\end{figure}

\begin{figure}[h]
 \centering
 \includegraphics[width=0.65\linewidth]{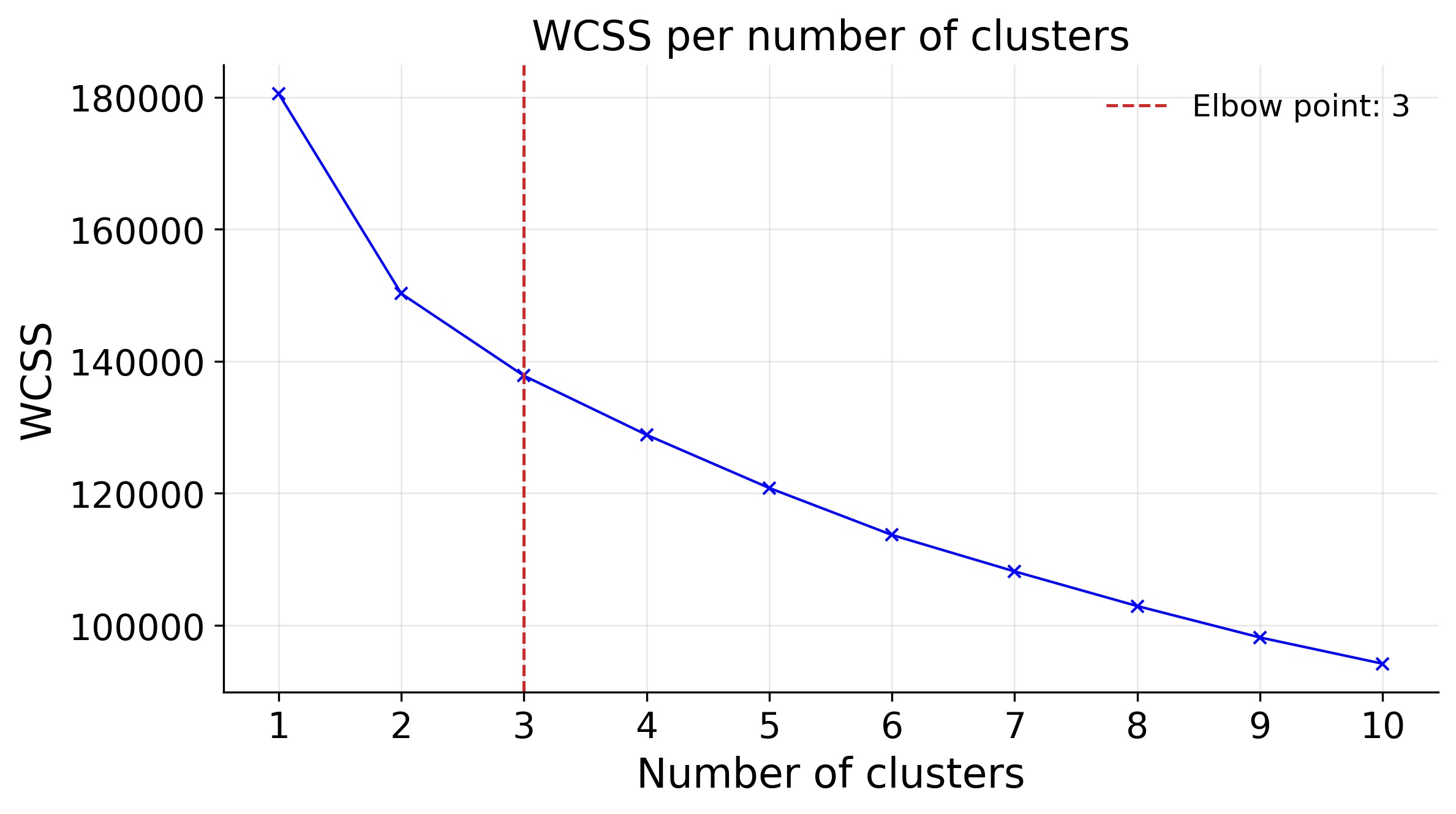}
 \caption{Within-Cluster Sum of Squares (WCSS) for the control group. The elbow is observed at \(K=3\).}
 \label{fig:control_wcss}
\end{figure}

\clearpage
\newpage

\section{Non-linear 2D UMAP of experimental groups' calls}
\label{sec:umap_supplementary}

\begin{figure}[h]
\centering
\includegraphics[width=0.95\textwidth]{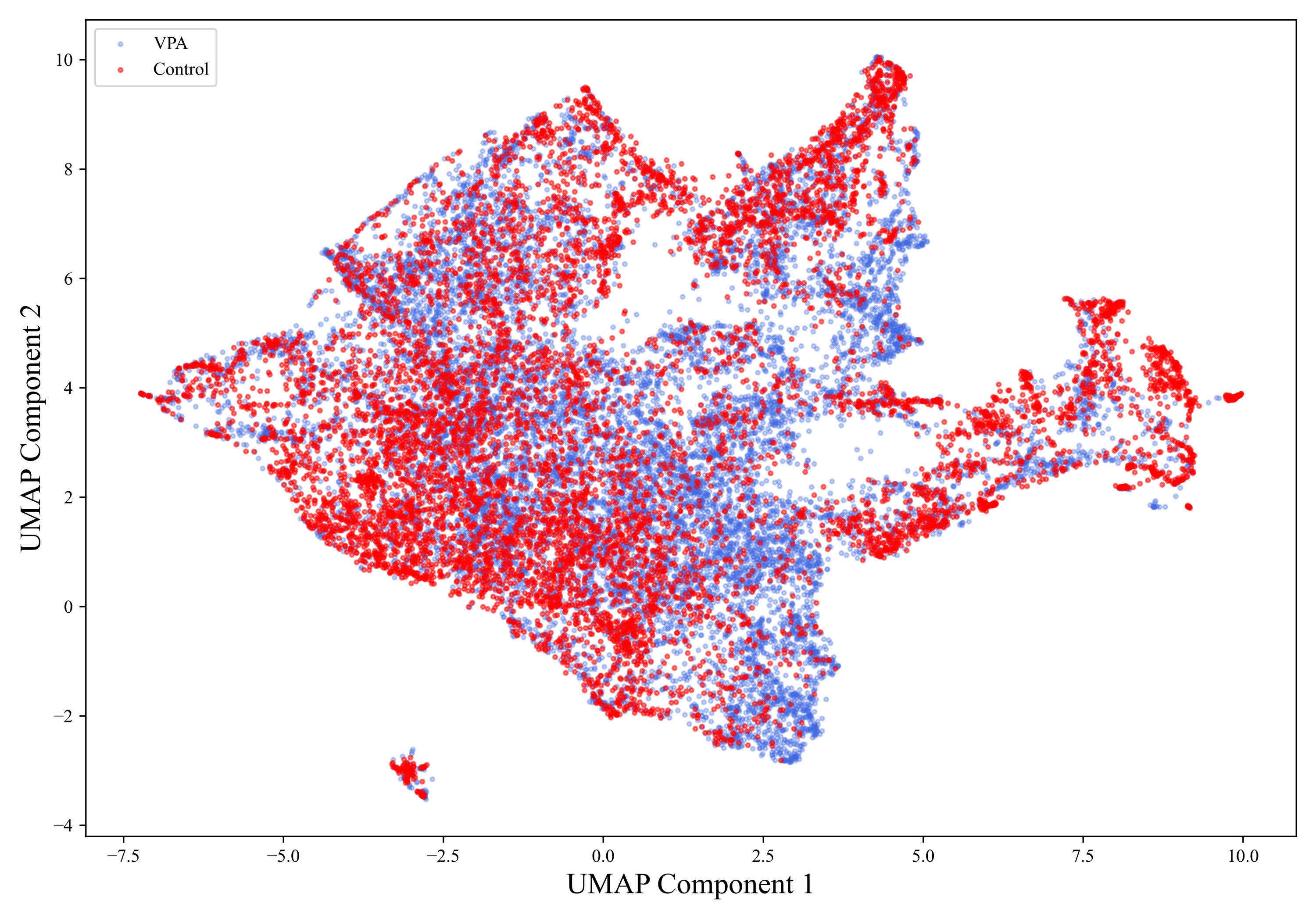}
\caption{2D UMAP of call-level acoustic features for control (red) and VPA (blue). UMAP was run on z-scored features (Euclidean metric)}
\label{fig:umap_2d_experimental_group}
\end{figure}

To visualise group structure without imposing linear assumptions, we embedded the 20 acoustic features using Uniform Manifold Approximation and Projection (UMAP). The 2D UMAP projection (Figure~\ref{fig:umap_2d_experimental_group}) offers a qualitative view of local neighbourhoods in the feature space. In this space, a substantial overlap is observed between control (red) and VPA (blue) calls, indicating largely shared variance across the feature set, with only localised regions of higher group concentration. 
This pattern suggests limited global separation in the manifold; given that UMAP preserves local structure and depends on hyperparameters and random initialisation, these projections should be interpreted as qualitative rather than inferential. Quantitative analysis for condition effects was therefore assessed at the feature level using linear mixed-effects models, as seen in Section~\ref{sec:lmm_main_section}.

\section{Linear Mixed Model for the number of calls}
\label{sec:lmm_calls_supplementary}

Table~\ref{tab:lmm_calls} reports the Linear Mixed Model results for the number of calls per time bin, with partial eta-squared ($\eta^2_p$) estimated from Type III F tests \cite{lakens2013calculating} with the \texttt{effectsize} package \cite{ben2020effectsize}. In the minimum adequate model, Cluster membership (χ² = 307.37, p < 0.001), Condition (χ² = 25.05, p < 0.001), their interaction (χ² = 23.01, p < 0.001) and Time bin (χ² = 22.95, p < 0.001) were significant, with more calls in time bins 5 and 6 than in time bins 1 and 2.

\begin{table}[h]
\centering
\scriptsize
\renewcommand{\arraystretch}{1.7}
\setlength{\tabcolsep}{3pt}
\resizebox{\textwidth}{!}{%
\begin{tabular}{>{\raggedright\arraybackslash}p{2.2cm}|ccc|ccc|ccc|ccc|ccc}
\hline
 & \multicolumn{3}{c|}{\textbf{Cluster}} & \multicolumn{3}{c|}{\textbf{Condition}} & \multicolumn{3}{c|}{\textbf{Temporal Bin}} & \multicolumn{3}{c|}{\textbf{Condition $\times$ Cluster}} & \multicolumn{3}{c}{\textbf{Condition $\times$ Cluster $\times$ Bin}} \\
& $\chi^2$ & $p$ & $\eta^2_p$ & $\chi^2$ & $p$ & $\eta^2_p$ & $\chi^2$ & $p$ & $\eta^2_p$ & $\chi^2$ & $p$ & $\eta^2_p$ & $\chi^2$ & $p$ & $\eta^2_p$ \\
\hline
Number of calls & 309.26 & $<0.001$ & 0.546 & 28.46 & 0.005 & 0.006 & 28.31 & 0.102 & 0.056 & 23.54 & $<0.001$ & 0.058 & 0.95 & 0.966 & 0.003 \\
\hline
\end{tabular}}
\caption{Linear Mixed Model results for the number of calls per time bin. Partial eta-squared ($\eta^2_p$) and statistical significance are reported for main effects and interactions in the full model.}
\label{tab:lmm_calls}
\end{table}

\section{Multicollinearity analysis}

\label{sec:multicollinearity_supplementary}

Before fitting the Linear Mixed Models, we assessed multicollinearity to avoid inflated variances and redundancy among predictors. 

Variance Inflation Factor (VIF) values were calculated for all acoustic features to assess the degree of multicollinearity in the dataset. Features that exceeded the conventional threshold of 10 were flagged as potentially problematic. The analysis revealed significant multicollinearity among several frequency domain features, with F0 Standard Deviation and F0 Bandwidth exhibiting the highest VIF values, at 28.77 and 28.67, respectively. This indicates that these features share approximately 96\% of their variance with other predictors in the model. Other features with elevated VIF values included Attack Magnitude (VIF = 8.27), RMS Standard Deviation (VIF = 7.97), RMS Mean (VIF = 7.49), F0 Mean (VIF = 5.48), and Spectral Centroid Mean (VIF = 4.65). 

\begin{figure}[ht]
 \centering
 \includegraphics[width=\textwidth]{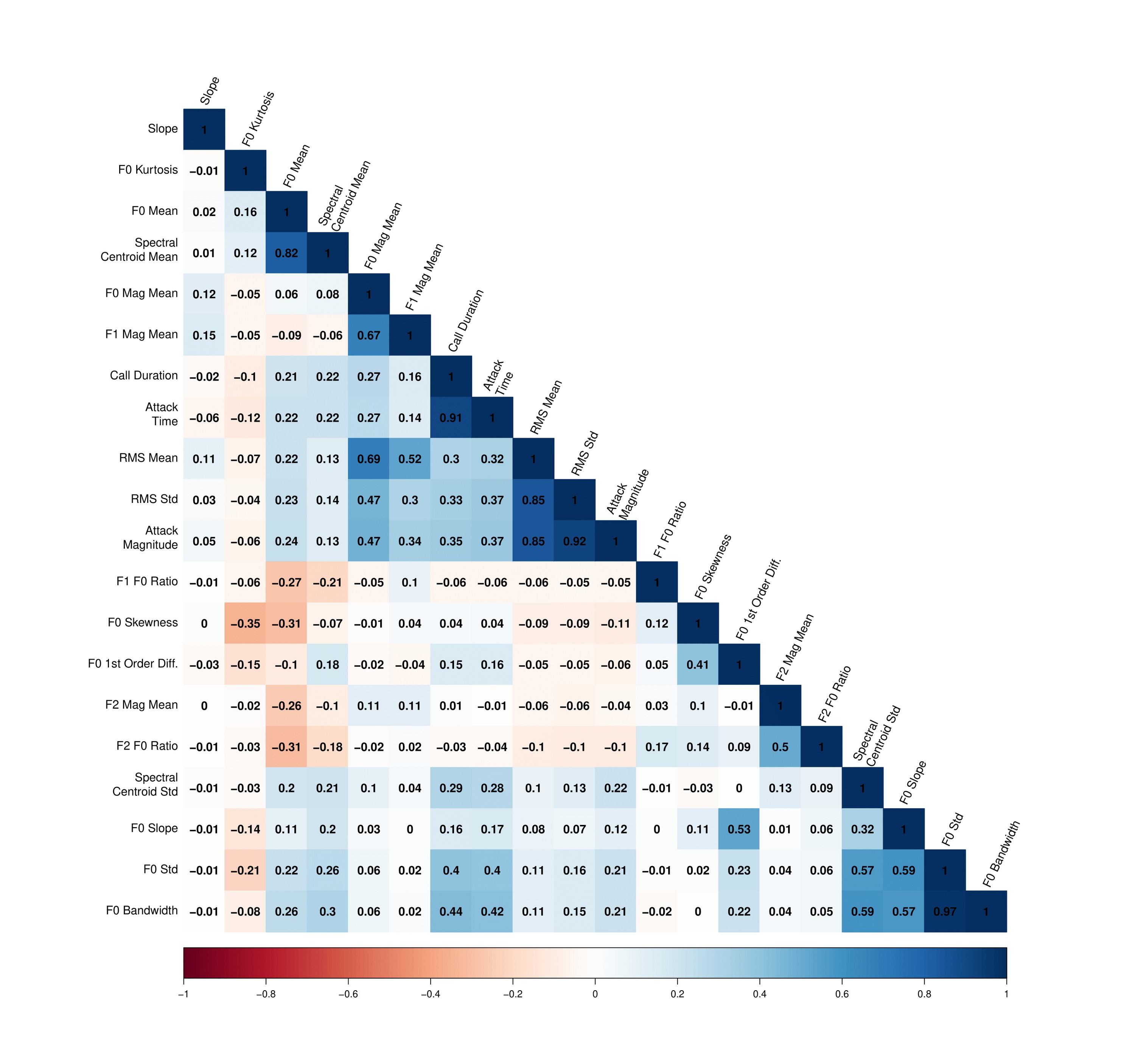}
\caption{Correlation matrix of acoustic features computed using Pearson’s correlation coefficients. Feature pairs exhibiting strong linear associations (|r| ≥ 0.8) were identified to assess multicollinearity and inform subsequent feature selection.}
 \label{fig:correlation_matrix_values}
\end{figure}

The pairwise correlation analysis revealed several highly correlated feature pairs (|r| ≥ 0.8), especially within the frequency and energy domains, as illustrated in Figure~\ref{fig:correlation_matrix_values}.

Notable high correlations included F0 Standard Deviation and F0 Bandwidth (r = 0.97), Attack time and Duration call (r = 0.91), RMS Standard Deviation and Attack magnitude (r = 0.92), RMS Mean and Attack magnitude (r = 0.85), RMS Mean and RMS Standard Deviation (r = 0.85), and F0 Mean and Spectral Centroid Mean (r = 0.82).

To address multicollinearity while preserving discriminative power, a systematic feature selection approach was implemented.
For each pair of highly correlated features, discriminative power was quantified using Cohen's d effect sizes for group differences. From the highly correlated pairs, the following features were selected based on superior discriminative power and interpretability: Spectral Centroid Mean (d = 0.364) over F0 Mean (d = 0.271); F0 Standard Deviation (d = 0.331) over F0 Bandwidth (d = 0.282); Call Duration (d = 0.175) over Attack Time (d = 0.208); and RMS Mean (d = 0.364) over Attack Magnitude (d = 0.385) and RMS Standard Deviation (d = 0.316).

In the last two cases, biological interpretability was prioritised over a slightly higher discriminative power: Call Duration and RMS Mean are directly associated with arousal in domestic chicks \cite{collins2024sound}, and RMS Mean is also directly comparable with the overall call amplitude previously reported in VPA-exposed chicks \cite{nishigori2013impaired}, whereas Attack Time and Attack Magnitude describe the onset transient, whose relationship to affective state is less directly documented.
The final feature set for the LMM analysis included 15 features: Duration Call, F0 Standard Deviation, F0 Skewness, F0 Kurtosis, F0 1st Order Difference, F0 Slope, F0 Magnitude Mean, F1 Magnitude Mean, F2 Magnitude Mean, F1/F0 Ratio Mean, F2/F0 Ratio Mean, Spectral Centroid Mean, Spectral Centroid Standard Deviation, RMS Mean and Envelope Slope. This feature selection ensured that the subsequent analyses were conducted with a robust set of predictors, covering the vocal acoustic space while maximising discriminatory power between experimental conditions.

\section{Feature analysis}
\subsection{Linear Mixed Models results across the 20 acoustic features}

This section reports the full results of the Linear Mixed Models (LMM) fitted on all 20 extracted acoustic features, analysed across six recording time bins and stratified by cluster and experimental condition (CTRL vs VPA). The models also included chick ID as a random intercept to account for repeated measures. Results are based on the 2-cluster solution identified via hierarchical clustering and are grouped by acoustic domain (temporal, energy, frequency).
Table~\ref{tab:lmm_complete} provides detailed statistics for all main effects and interactions, including $\chi^2$, FDR-corrected $p$-values \cite{benjamini1995controlling}, and $\eta^2_p$ effect sizes computed from Type III F tests \cite{lakens2013calculating}, using the \texttt{effectsize} package \cite{ben2020effectsize}. Partial eta-squared values of 0.01, 0.06 and 0.14 indicate, respectively, small, medium and large effects \cite{cohen2013statistical}.
Large $\chi^2$ values with small $\eta^2_p$ values indicate effects that depend on cluster or time bin rather than uniform effects. The 15 features retained in the main analysis are discussed in the Results section \ref{sec:lmm_main_section}; additional features are included here for completeness.
\label{sec:lmm_supplementary}

\begin{landscape}
\begin{table}[p]
\centering
\scriptsize
\renewcommand{\arraystretch}{1.7}
\setlength{\tabcolsep}{3pt}
\begin{tabular}{>{\raggedright\arraybackslash}p{2.8cm}|ccc|ccc|ccc|ccc|ccc}
\hline
\multirow{2}{*}{\textbf{Feature}} & \multicolumn{3}{c|}{\textbf{Cluster}} & \multicolumn{3}{c|}{\textbf{Condition}} & \multicolumn{3}{c|}{\textbf{Temporal Bin}} & \multicolumn{3}{c|}{\textbf{Condition $\times$ Cluster}} & \multicolumn{3}{c}{\textbf{Condition $\times$ Cluster $\times$ Bin}} \\
& $\chi^2$ & $p$ & $\eta^2_p$ & $\chi^2$ & $p$ & $\eta^2_p$ & $\chi^2$ & $p$ & $\eta^2_p$ & $\chi^2$ & $p$ & $\eta^2_p$ & $\chi^2$ & $p$ & $\eta^2_p$ \\
\hline
\multicolumn{16}{l}{\textbf{Temporal Domain}} \\
\hline
Call Duration & 3846.09 & $<0.001$ & 0.077 & 629.82 & $<0.001$ & 0.467 & 413.08 & $<0.001$ & 0.002 & 617.35 & $<0.001$ & 0.016 & 54.66 & $<0.001$ & 0.002 \\
Attack Time & 3896.02 & $<0.001$ & 0.079 & 616.08 & $<0.001$ & 0.469 & 345.69 & $<0.001$ & 0.003 & 595.90 & $<0.001$ & 0.015 & 44.41 & $<0.001$ & 0.002 \\
Slope & 282.87 & $<0.001$ & 0.017 & 71.89 & $<0.001$ & $<0.001$ & 205.01 & $<0.001$ & 0.003 & 62.61 & $<0.001$ & $<0.001$ & 62.60 & $<0.001$ & 0.003 \\
\hline
\multicolumn{16}{l}{\textbf{Energy Domain}} \\
\hline
RMS Mean & 14119.73 & $<0.001$ & 0.342 & 1013.94 & $<0.001$ & 0.019 & 2605.85 & $<0.001$ & 0.043 & 730.42 & $<0.001$ & 0.011 & 471.56 & $<0.001$ & 0.020 \\
RMS Standard Deviation & 9294.08 & $<0.001$ & 0.248 & 543.23 & $<0.001$ & 0.054 & 1769.53 & $<0.001$ & 0.028 & 388.98 & $<0.001$ & 0.013 & 79.76 & $<0.001$ & 0.003 \\
Attack Magnitude & 10029.14 & $<0.001$ & 0.276 & 688.62 & $<0.001$ & 0.061 & 1542.17 & $<0.001$ & 0.020 & 556.52 & $<0.001$ & 0.020 & 139.37 & $<0.001$ & 0.006 \\
\hline
\multicolumn{16}{l}{\textbf{Frequency Domain}} \\
\hline
F0 Mean & 1108.25 & $<0.001$ & 0.035 & 107.27 & $<0.001$ & 0.017 & 469.74 & $<0.001$ & 0.001 & 50.68 & $<0.001$ & $<0.001$ & 11.49 & 0.048 & $<0.001$ \\
F0 Standard Deviation & 3850.06 & $<0.001$ & 0.080 & 1013.91 & $<0.001$ & 0.528 & 463.79 & $<0.001$ & 0.006 & 927.37 & $<0.001$ & 0.033 & 57.12 & $<0.001$ & 0.002 \\
F0 Skewness & 31.15 & 0.002 & $<0.001$ & 21.64 & 0.048 & $<0.001$ & 105.29 & $<0.001$ & $<0.001$ & 11.34 & 0.086 & $<0.001$ & 11.29 & 0.051 & $<0.001$ \\
F0 Kurtosis & 260.25 & $<0.001$ & 0.007 & 54.72 & $<0.001$ & 0.254 & 123.76 & $<0.001$ & $<0.001$ & 27.72 & $<0.001$ & $<0.001$ & 17.05 & 0.005 & $<0.001$ \\
F0 Bandwidth & 3417.33 & $<0.001$ & 0.072 & 861.09 & $<0.001$ & 0.452 & 418.59 & $<0.001$ & 0.005 & 773.95 & $<0.001$ & 0.027 & 47.42 & $<0.001$ & 0.002 \\
F0 1st Order Difference & 921.62 & $<0.001$ & 0.020 & 184.11 & $<0.001$ & 0.125 & 290.95 & $<0.001$ & 0.003 & 170.88 & $<0.001$ & 0.007 & 42.12 & $<0.001$ & 0.002 \\
F0 Slope & 3467.01 & $<0.001$ & 0.076 & 766.67 & $<0.001$ & 0.521 & 988.87 & $<0.001$ & 0.016 & 677.26 & $<0.001$ & 0.025 & 100.38 & $<0.001$ & 0.004 \\
F0 Mag Mean & 5725.45 & $<0.001$ & 0.104 & 825.49 & $<0.001$ & 0.211 & 1584.43 & $<0.001$ & 0.027 & 678.42 & $<0.001$ & 0.007 & 344.90 & $<0.001$ & 0.015 \\
F1 Mag Mean & 3273.14 & $<0.001$ & 0.064 & 234.11 & $<0.001$ & 0.002 & 841.67 & $<0.001$ & 0.016 & 166.81 & $<0.001$ & $<0.001$ & 164.00 & $<0.001$ & 0.007 \\
F2 Mag Mean & 43.37 & $<0.001$ & 0.001 & 19.66 & 0.082 & 0.003 & 299.11 & $<0.001$ & $<0.001$ & 7.29 & 0.315 & $<0.001$ & 1.24 & 0.951 & $<0.001$ \\
F1/F0 Ratio Mean & 63.27 & $<0.001$ & 0.002 & 26.08 & 0.012 & 0.016 & 53.13 & $<0.001$ & $<0.001$ & 14.93 & 0.024 & $<0.001$ & 4.89 & 0.449 & $<0.001$ \\
F2/F0 Ratio Mean & 47.68 & $<0.001$ & 0.002 & 16.41 & 0.188 & 0.045 & 122.86 & $<0.001$ & $<0.001$ & 2.93 & 0.836 & $<0.001$ & 1.21 & 0.951 & $<0.001$ \\
Spectral Centroid Mean & 1605.07 & $<0.001$ & 0.046 & 166.94 & $<0.001$ & 0.024 & 267.72 & $<0.001$ & 0.001 & 99.57 & $<0.001$ & 0.003 & 9.50 & 0.099 & $<0.001$ \\
Spectral Centroid Std & 866.20 & $<0.001$ & 0.021 & 251.38 & $<0.001$ & 0.287 & 324.61 & $<0.001$ & 0.002 & 105.14 & $<0.001$ & 0.003 & 17.77 & 0.004 & $<0.001$ \\
\hline
\end{tabular}
\caption{Linear Mixed Models results for acoustic features across temporal, energy, and frequency domains. Partial eta-squared ($\eta^2_p$) and statistical significance are reported for main effects and interactions following FDR correction ($\alpha = 0.05$).}
\label{tab:lmm_complete}
\end{table}
\end{landscape}

\clearpage
\newpage
\subsection{Complete statistical results: additional significant features}
\label{section:plot_significant_collinear}
This section reports detailed results and plots of features that showed significant effects but were excluded from the main text due to lower effect sizes or high correlations with other features. 
Although omitted for brevity, these results support the robustness of the main findings.

\subsubsection{Temporal domain features.}
\label{sec:Temporal_domain_suppl_plots}

We found similar statistics for Attack Time and Envelope Slope, with the minimum adequate model including the interaction between Condition, Cluster membership, and Time bin (see Supplementary Table~\ref{tab:lmm_complete}).

\begin{figure}[!htbp]
 \centering
 \includegraphics[width=1\textwidth]{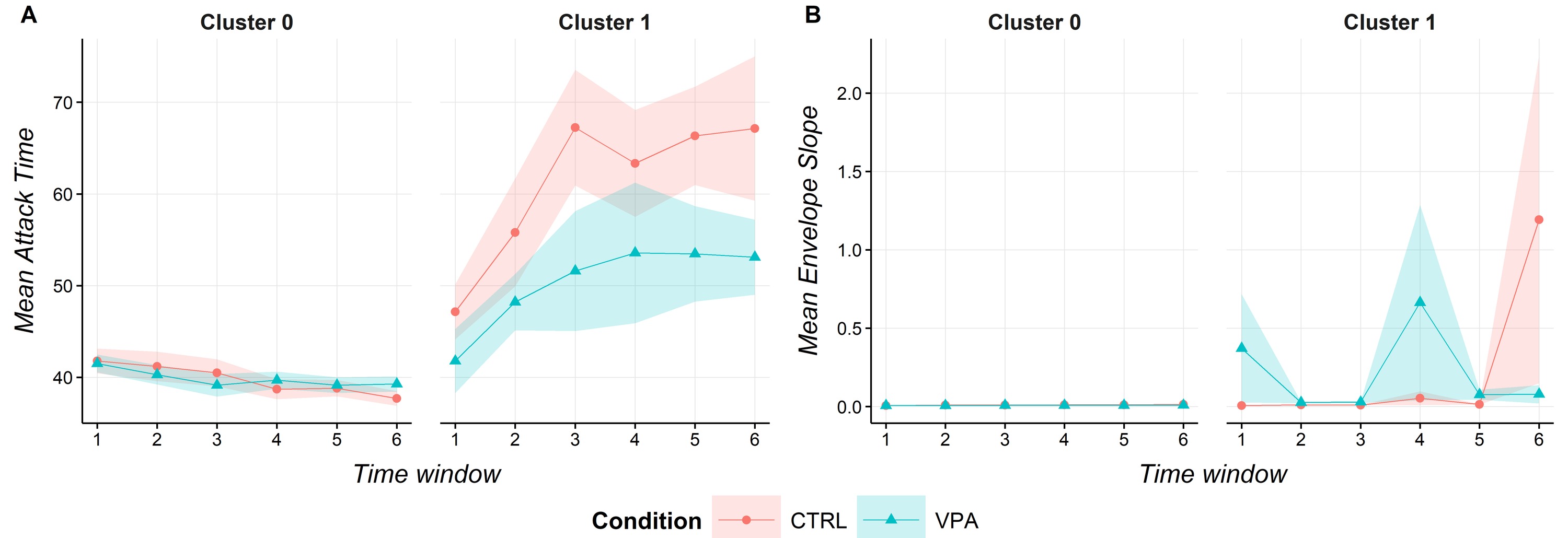}
 \caption{Temporal-domain features across time bins by condition and cluster. 
 (A) Attack Time, (B) Envelope Slope. Mean ± SEM}
 \label{fig:attack_slope_complete}
\end{figure}

For Attack Time, the analysis revealed a significant three-way interaction (χ² = 44.41, p < 0.001, $\eta^2_p$ = 0.002) and a condition × cluster interaction (χ² = 595.90, p < 0.001, $\eta^2_p$ = 0.015). This feature was excluded from the main analysis due to high correlation with Call Duration (r = 0.91). As expected, the pattern of results mirrors those observed for Call Duration: while cluster 0 calls showed no differences between conditions, cluster 1 calls exhibited significantly longer attack times in control compared to VPA chicks (post-hoc: p < 0.001). This difference remained stable across the recording session (see Figure~\ref{fig:attack_slope_complete}A).

For Envelope Slope, although included in the main analysis, its temporal profile was more variable and less readily interpretable than that of Call Duration; its results are therefore reported here. The analysis revealed a significant three-way interaction (χ² = 62.60, p < 0.001, $\eta^2_p$ = 0.003) and a significant condition × cluster interaction (χ² = 62.61, p < 0.001, $\eta^2_p$ < 0.001). Cluster 1 calls demonstrated condition-specific temporal dynamics that varied across the recording session (see Figure~\ref{fig:attack_slope_complete}B).

% \newpage
\subsubsection{Energy domain features.}
\label{sec:energy_domain_suppl_plots}

\begin{figure}[!htbp]
 \centering
 \includegraphics[width=1\textwidth]{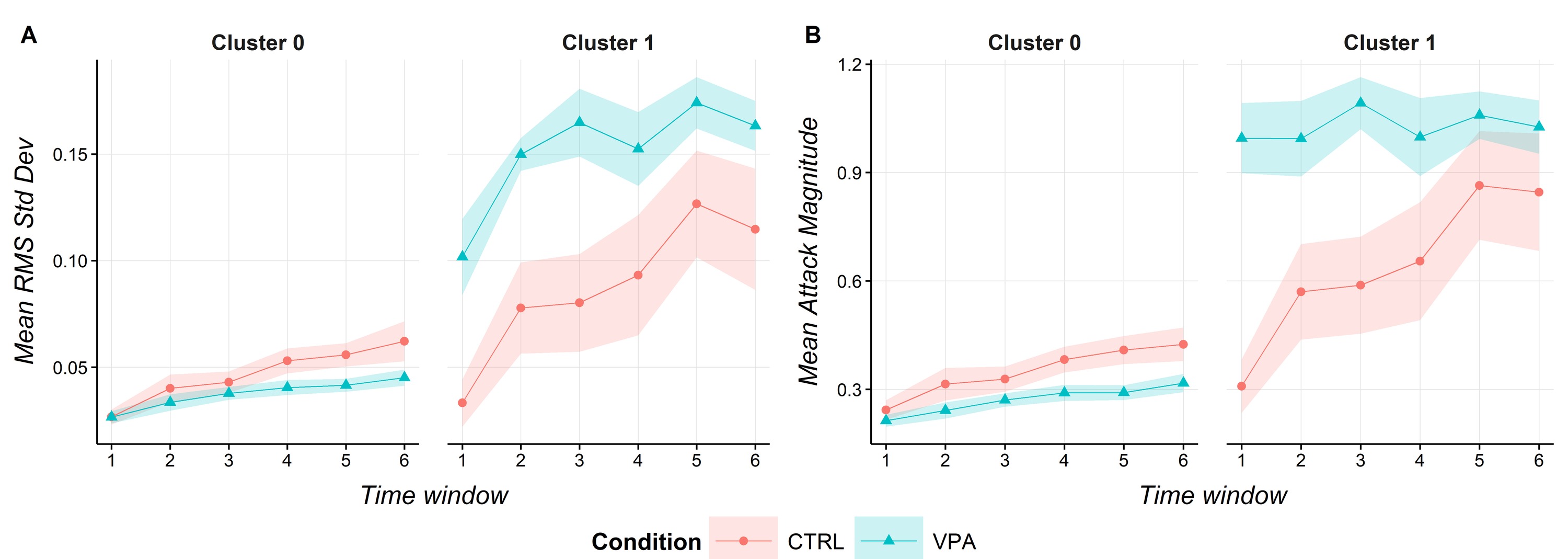}
 \caption{Energy-domain features across time bins by condition and cluster. 
 (A) RMS Standard Deviation, (B) Attack Magnitude. Mean ± SEM}
 \label{fig:Energy_related_features_suppl}
\end{figure}

For RMS Standard Deviation, the minimum adequate model included the interaction between Condition, Cluster membership, and Time bin (see Supplementary Table~\ref{tab:lmm_complete}). 
The analysis revealed significant main effects for Cluster (χ² = 9294.08, p < 0.001, $\eta^2_p$ = 0.248), Condition (χ² = 543.23, p < 0.001, $\eta^2_p$ = 0.054), and Temporal Bin (χ² = 1769.53, p < 0.001, $\eta^2_p$ = 0.028), as well as a condition × cluster interaction (χ² = 388.98, p < 0.001, $\eta^2_p$ = 0.013). This feature was excluded from the main text due to high correlation with RMS Mean (r = 0.85), though it displays a complementary pattern. Cluster 1 calls showed higher energy variability in VPA-exposed chicks compared to controls (post-hoc: p < 0.001). In contrast, cluster 0 calls exhibited slightly reduced energy variability under VPA conditions, although this difference was not significant (post-hoc: p = 0.151). These findings support that VPA exposure affects both the average energy level and its variability in a call-type-specific manner (Figure~\ref{fig:Energy_related_features_suppl}).

Finally, for Attack Magnitude, the minimum adequate model included the interaction between Condition, Cluster membership, and Time bin (see Supplementary Table~\ref{tab:lmm_complete}). The analysis revealed significant main effects for Cluster (χ² = 10029.14, p < 0.001, $\eta^2_p$ = 0.276), Condition (χ² = 688.62, p < 0.001, $\eta^2_p$ = 0.061), and Temporal Bin (χ² = 1542.17, p < 0.001, $\eta^2_p$ = 0.020), as well as a condition × cluster interaction (χ² = 556.52, p < 0.001, $\eta^2_p$ = 0.020). This feature was also excluded from the main text due to its high correlation with RMS Standard Deviation (r = 0.92) and RMS Mean (r = 0.85). Cluster 1 calls showed higher attack magnitude values in VPA-exposed chicks compared to controls (post-hoc: p < 0.001), whereas cluster 0 calls exhibited the opposite pattern, with lower values under VPA conditions (post-hoc: p = 0.031). Notably, while control chicks demonstrated progressive increases in attack magnitude across the recording session, VPA-exposed chicks maintained consistently high values with reduced temporal variability (Figure~\ref{fig:Energy_related_features_suppl}B).

\subsubsection{Frequency domain features.}
\label{sec:frequency_domain_suppl_plots}

We found similar patterns for features related to fundamental frequency (F0 Mean, F0 Skewness, F0 Magnitude Mean), with the minimum adequate model including the interaction between Condition, Cluster membership, and Time bin (see Supplementary Table~\ref{tab:lmm_complete}).

\begin{figure}[!htbp]
 \centering
 \includegraphics[width=\textwidth]{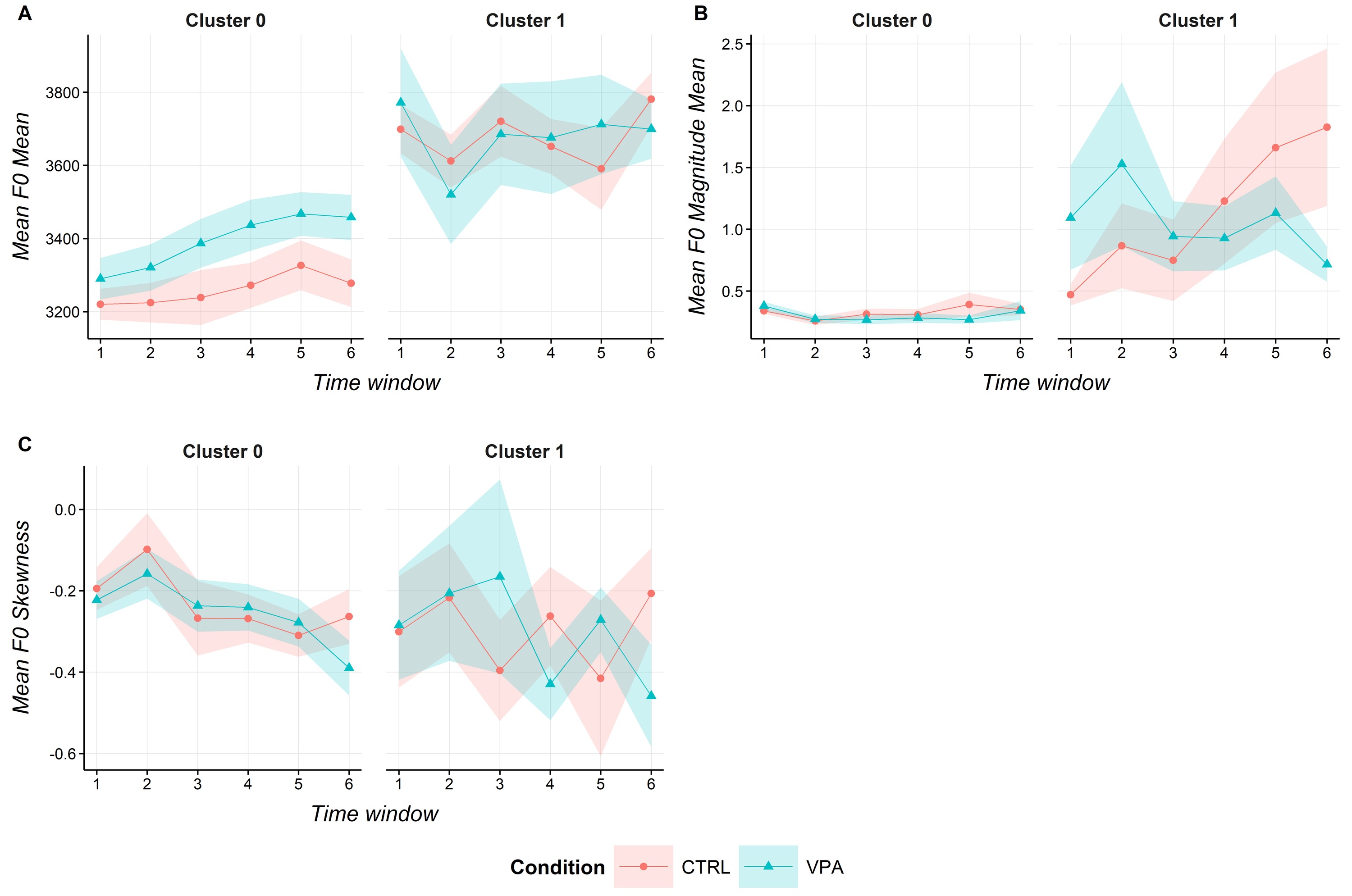}
 \caption{Frequency-domain features across 6 time bins by condition and cluster. 
 (A) F0 Mean, (B) F0 Magnitude Mean, (C) F0 Skewness. Mean ± SEM}
 \label{fig:frequency_features_suppl}
\end{figure}

For F0 Mean, the analysis showed significant main effects for Cluster (χ² = 1108.25, p < 0.001, $\eta^2_p$ = 0.035) and Condition (χ² = 107.27, p < 0.001, $\eta^2_p$ = 0.017), with a small three-way interaction (χ² = 11.49, p = 0.048, $\eta^2_p$ < 0.001). This feature was excluded from the main analysis due to high correlation with Spectral Centroid Mean (r = 0.82). While post-hoc comparisons did not reveal significant condition differences within clusters (cluster 0: p = 0.156; cluster 1: p = 0.717), temporal dynamics suggested subtle condition-dependent variations in fundamental frequency that varied during the recording session (Figure~\ref{fig:frequency_features_suppl}A).
For F0 Magnitude Mean, the analysis showed a significant three-way interaction (χ² = 344.90, p < 0.001, $\eta^2_p$ = 0.015). This feature exhibited a striking temporal inversion pattern, particularly for cluster 1 calls. VPA-exposed chicks initially showed higher F0 magnitude values, but these progressively declined over the recording session. In contrast, control chicks demonstrated the opposite temporal trajectory (post-hoc cluster 1: p < 0.001). Cluster 0 calls showed no significant condition differences (post-hoc: p = 0.908) (see Figure~\ref{fig:frequency_features_suppl}B).

For F0 Skewness, the minimum adequate model included the three-way interaction, which did not remain significant after FDR correction (χ² = 11.29, p = 0.051, $\eta^2_p$ < 0.001). The effect sizes were also considerably smaller than those observed for other frequency variability features reported in the main text (F0 Standard Deviation, F0 Kurtosis). Post-hoc comparisons showed no significant differences within clusters (cluster 0: p = 0.970; cluster 1: p = 0.959), confirming that while F0 Skewness differed between call types and showed subtle condition effects, these were less pronounced than other pitch distribution measures (Figure~\ref{fig:frequency_features_suppl}C).

Harmonic-related features revealed distinctive patterns based on their acoustic properties. 
We examined both magnitude features (F1 Magnitude Mean, F2 Magnitude Mean) and ratio features (F1/F0 Ratio Mean, F2/F0 Ratio Mean).

\begin{figure}[!htbp]
 \centering
 \includegraphics[width=\textwidth]{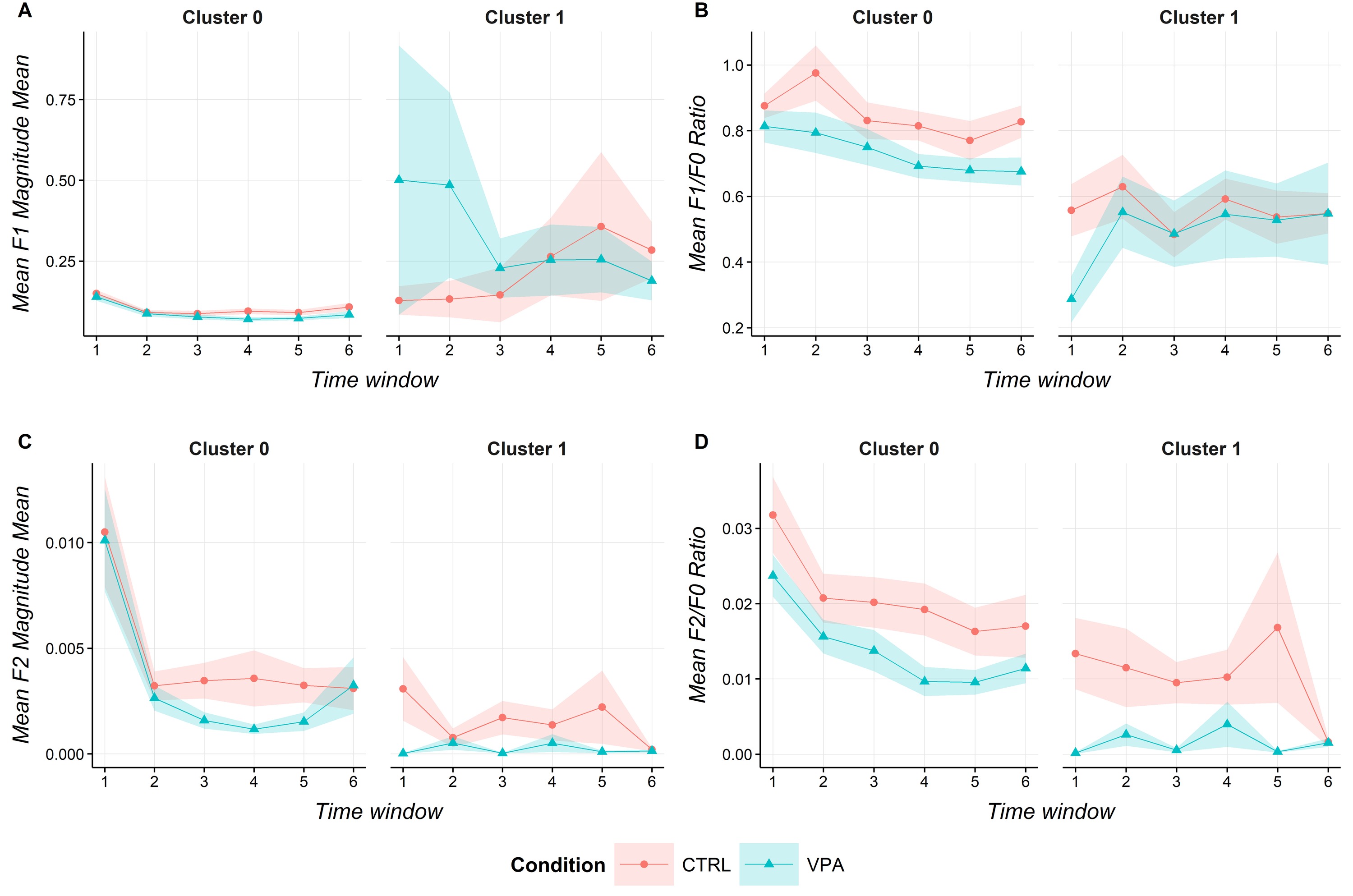}
 \caption{Harmonic-related features across 6 time bins by condition and cluster. 
 (A) F1 Magnitude Mean, (B) F1–F0 Ratio, (C) F2 Magnitude Mean, (D) F2–F0 Ratio. Mean ± SEM.}
 \label{fig:freq_features_formants}
\end{figure}

For F1 Magnitude Mean, the minimum adequate model included the interaction between Condition, Cluster membership, and Time bin (see Supplementary Table~\ref{tab:lmm_complete}), with a significant three-way interaction (χ² = 164.00, p < 0.001, $\eta^2_p$ = 0.007). Similar to F0 Magnitude Mean, this feature showed temporal inversion patterns. VPA-treated chicks showed a progressive reduction in first formant magnitude across later time bins, while control values increased over time. However, post-hoc comparisons revealed no significant condition differences within either cluster (cluster 0: p = 0.743; cluster 1: p = 0.811), likely due to high within-group variability and the complex temporal dynamics (Figure~\ref{fig:freq_features_formants}A).

For F1/F0 Ratio Mean, the minimum adequate model included the condition × cluster interaction but not the three-way interaction (see Supplementary Table~\ref{tab:lmm_complete}; χ² = 4.89, p = 0.449). The analysis revealed significant main effects of Cluster (χ² = 63.27, p < 0.001, $\eta^2_p$ = 0.002), Condition (χ² = 26.08, p = 0.012, $\eta^2_p$ = 0.016) and Temporal Bin (χ² = 53.13, p < 0.001, $\eta^2_p$ < 0.001), together with a weak but significant condition × cluster interaction (χ² = 14.93, p = 0.024, $\eta^2_p$ < 0.001). The pattern differed markedly from magnitude features. Cluster 0 calls showed a small but consistent divergence between conditions across time bins (post-hoc: p = 0.033), with a lower F1/F0 ratio Mean in VPA-exposed chicks, indicating a weaker first harmonic relative to the fundamental. In contrast, cluster 1 calls did not differ between conditions, displaying large within-group variability and overlapping trends, particularly after 120 seconds (post-hoc: p = 0.421). Unlike F1 Magnitude Mean, the F1/F0 Ratio Mean showed a significant effect in cluster 0, reflecting a difference in spectral balance not visible in the absolute magnitude (Figure~\ref{fig:freq_features_formants}B).

For F2 Magnitude Mean, the three-way interaction was not significant (see Supplementary Table~\ref{tab:lmm_complete}; χ² = 1.24, p = 0.951). The analysis showed a significant main effect of Cluster (χ² = 43.37, p < 0.001, $\eta^2_p$ = 0.001), while the condition × cluster interaction did not reach significance (χ² = 7.29, p = 0.315, $\eta^2_p$ < 0.001). Unlike F0 and F1 magnitude features, F2 Magnitude Mean showed more stable patterns across time bins. Post-hoc comparisons suggested subtle differences between conditions (cluster 0: p = 0.077; cluster 1: p = 0.286), but the effects were less consistent and of smaller magnitude compared to the lower formants (Figure~\ref{fig:freq_features_formants}C).

Finally, for F2/F0 Ratio Mean, the minimum adequate model included only main effects (see Supplementary Table~\ref{tab:lmm_complete}), as neither the three-way interaction (χ² = 1.21, p = 0.951) nor the condition × cluster interaction (χ² = 2.93, p = 0.836) were significant. The analysis showed a significant main effect of Cluster (χ² = 47.68, p < 0.001, $\eta^2_p$ = 0.002), while the main effect of Condition was not significant (χ² = 16.41, p = 0.188, $\eta^2_p$ = 0.045). Overall, F2/F0 Ratio Mean values were lower in VPA-exposed chicks, but this difference did not reach significance, likely due to the high within-group variability of this feature (Figure~\ref{fig:freq_features_formants}D).

\clearpage
\newpage
\subsection{Post Hoc Tukey statistical results}
\label{section:post_hoc}

To further characterise the significant \textit{condition} $\times$ \textit{cluster} interactions identified in the LMM analysis, we conducted post-hoc Tukey-adjusted pairwise comparisons using the \texttt{emmeans} package \cite{lenth2024emmeans}. Table~\ref{tab:posthoc_by_cluster} presents all acoustic features that showed significant differences between VPA-exposed and control groups within specific clusters, organised by acoustic domain. The results reveal that cluster 1 calls exhibited substantially more condition-sensitive features across all acoustic domains compared to cluster 0, indicating differential vulnerability to prenatal VPA exposure between call types.

\begin{table}[h]
\centering
\footnotesize
\renewcommand{\arraystretch}{0.85}
\begin{tabular}{>{\centering\arraybackslash}p{5.5cm}|>{\centering\arraybackslash}p{5.5cm}}
\toprule
\textbf{Cluster 0 } & \textbf{Cluster 1 } \\
\midrule
\multicolumn{2}{c}{\textbf{Temporal Domain}} \\
\midrule
 & \textit{Call Duration} \\
 & CTRL > VPA ($p<0.001$) \\
 & \\
 & \textit{Attack Time} \\
 & CTRL > VPA ($p<0.001$) \\
\midrule
\multicolumn{2}{c}{\textbf{Energy Domain}} \\
\midrule
& \textit{ RMS Mean} \\
 & CTRL < VPA ($p=0.001$) \\
 & \\
 & \textit{RMS Standard Deviation} \\
 & CTRL < VPA ($p<0.001$) \\
 & \\
 \textit{Attack Magnitude} & \textit{Attack Magnitude} \\
CTRL > VPA ($p=0.031$) & CTRL < VPA ($p<0.001$) \\
\midrule
\multicolumn{2}{c}{\textbf{Frequency Domain}} \\
\midrule
\textit{Spectral Centroid Standard Deviation} & \textit{Spectral Centroid Standard Deviation} \\
CTRL > VPA ($p=0.005$) & CTRL > VPA ($p<0.001$) \\
 & \\
\textit{F0 Kurtosis} & \textit{F0 Kurtosis} \\
CTRL < VPA ($p<0.001$) & CTRL < VPA ($p<0.001$) \\
 & \\
 & \textit{F0 Standard Deviation} \\
 & CTRL > VPA ($p<0.001$) \\
 & \\
 & \textit{F0 Bandwidth} \\
 & CTRL > VPA ($p<0.001$) \\
 & \\
 & \textit{F0 1st Order Difference} \\
 & CTRL > VPA ($p<0.001$) \\
 & \\
 & \textit{F0 Slope} \\
 & CTRL > VPA ($p<0.001$) \\
 & \\
 & \textit{F0 Magnitude Mean} \\
 & CTRL > VPA ($p<0.001$) \\
 & \\
\textit{F1/F0 Ratio Mean} & \\
CTRL > VPA ($p=0.033$) & \\
\bottomrule
\end{tabular}
\vspace{0.5em}
\caption{Post-hoc Tukey test results for significant \textit{Condition} $\times$ \textit{Cluster} interactions. The table reports acoustic features significantly different between the VPA and CTRL groups within either cluster 0 or 1 (Tukey-adjusted). CTRL > VPA: control group had higher values than VPA-exposed; CTRL < VPA: control group had lower values.}
\label{tab:posthoc_by_cluster}
\end{table}

\newpage
\section{Average calls and comparison with literature}
\label{average_calls_and_literature_comparison}

The following section presents the mean values for acoustic features across clusters identified in this study for both the development dataset and the experimental datasets (control and VPA groups).

\begin{table}[htbp]
\centering
\scriptsize
\setlength{\tabcolsep}{4pt}
\renewcommand{\arraystretch}{1.1}
\begin{tabular}{lcccccc}
\toprule
& \multicolumn{2}{c}{Development data} & \multicolumn{2}{c}{Control group} & \multicolumn{2}{c}{VPA group}\\
\cmidrule(lr){2-3} \cmidrule(lr){4-5} \cmidrule(lr){6-7}
Feature & Cluster 0 & Cluster 1 & Cluster 0 & Cluster 1 & Cluster 0 & Cluster 1 \\
& (n=3499) & (n=2425) & (n=7661) & (n=1367) & (n=13632) & (n=889) \\
\midrule
Temporal Domain & & & & & & \\
Duration (s) & 0.142 ± 0.047 & 0.346 ± 0.089 & 0.115 ± 0.033 & 0.186 ± 0.058 & 0.117 ± 0.032 & 0.145 ± 0.045 \\
Attack Time & 26.40 ± 14.35 & 63.36 ± 16.98 & 39.47 ± 14.81 & 73.64 ± 29.04 & 40.05 ± 14.87 & 53.40 ± 21.55 \\
\midrule
Energy Domain & & & & & & \\
RMS Mean & 0.025 ± 0.018 & 0.138 ± 0.057 & 0.097 ± 0.090 & 0.383 ± 0.271 & 0.070 ± 0.052 & 0.420 ± 0.158 \\
RMS Std & 0.015 ± 0.016 & 0.130 ± 0.054 & 0.051 ± 0.048 & 0.155 ± 0.102 & 0.041 ± 0.030 & 0.177 ± 0.063 \\
Attack Magnitude & 0.133 ± 0.119 & 0.909 ± 0.290 & 0.382 ± 0.294 & 1.023 ± 0.576 & 0.290 ± 0.196 & 1.181 ± 0.325 \\
\midrule
Frequency Domain & & & & & & \\
F0 Mean (Hz) & 3022 ± 482 & 3332 ± 276 & 3272 ± 426 & 3782 ± 305 & 3471 ± 496 & 3582 ± 347 \\
F0 Std (Hz) & 282 ± 132 & 352 ± 143 & 173 ± 105 & 404 ± 200 & 163 ± 115 & 190 ± 131 \\
F0 Skewness & 0.07 ± 0.67 & 0.33 ± 0.66 & -0.29 ± 0.82 & -0.36 ± 0.58 & -0.34 ± 0.95 & -0.44 ± 0.66 \\
F0 Kurtosis & -0.78 ± 0.85 & -0.48 ± 0.77 & -0.58 ± 1.14 & -1.10 ± 0.88 & -0.25 ± 1.31 & -0.68 ± 1.16 \\
F0 Bandwidth (Hz) & 855 ± 381 & 1207 ± 419 & 502 ± 282 & 1073 ± 466 & 488 ± 314 & 551 ± 343 \\
F0 1st Order Diff & 37.02 ± 69.66 & 8.65 ± 32.82 & -2.88 ± 51.79 & 39.90 ± 68.67 & 3.93 ± 52.11 & 14.66 ± 41.14 \\
F0 Slope & 7041 ± 5454 & 11844 ± 8479 & 3408 ± 3905 & 11486 ± 11434 & 3174 ± 4359 & 4681 ± 4631 \\
F0 Magnitude Mean & 0.247 ± 0.509 & 0.614 ± 0.509 & 0.329 ± 0.562 & 2.508 ± 2.885 & 0.288 ± 0.418 & 1.673 ± 2.194 \\
F1 Magnitude Mean & 0.030 ± 0.056 & 0.011 ± 0.019 & 0.100 ± 0.108 & 0.467 ± 0.829 & 0.077 ± 0.089 & 0.487 ± 0.836 \\
F2 Magnitude Mean & 0.000070 ± 0.000 & 0.00000057 ± 0.000 & 0.0037 ± 0.019 & 0.00052 ± 0.003 & 0.0022 ± 0.015 & 0.00017 ± 0.001 \\
F1/F0 Ratio Mean & 0.282 ± 0.419 & 0.067 ± 0.105 & 0.828 ± 1.333 & 0.545 ± 0.486 & 0.669 ± 0.751 & 0.610 ± 0.569 \\
F2/F0 Ratio Mean & 0.00033 ± 0.003 & 0.0000051 ± 0.000 & 0.0187 ± 0.074 & 0.0043 ± 0.018 & 0.0109 ± 0.051 & 0.0012 ± 0.010 \\
Spectral Centroid Mean (Hz) & 5594 ± 1146 & 5835 ± 832 & 5946 ± 1188 & 7374 ± 925 & 6610 ± 1252 & 6723 ± 1038 \\
Spectral Centroid Std (Hz) & 1040 ± 413 & 1185 ± 400 & 908 ± 518 & 1370 ± 537 & 757 ± 452 & 798 ± 452 \\
Envelope Slope & 0.0060 ± 0.005 & 0.015 ± 0.006 & 0.0110 ± 0.027 & 0.113 ± 1.348 & 0.0080 ± 0.009 & 0.112 ± 0.956 \\
\bottomrule
\end{tabular}
\caption{Mean ± standard deviation of acoustic features for the identified clusters across development and experimental datasets (control and VPA groups).}
\label{tab:combined_cluster_characteristics}
\end{table}

The following results are presented in comparison to the findings of the existing literature. Table~\ref{tab:literature_computational_comparison} compares our computational framework results with the classifications reported by Marx et al. \cite{marx2001vocalisation}, highlighting quantitative differences.

\begin{landscape}
\begin{table}[htbp]
\centering
\footnotesize
\renewcommand{\arraystretch}{1.4}
\begin{tabular}{|p{2.7cm}|p{1.7cm}|p{1.3cm}|p{1.9cm}|p{1.7cm}|p{4.6cm}|p{6.2cm}|}
\toprule
\multirow{2}{*}{Dataset} & \multirow{2}{*}{ Cluster} & Duration & Energy & F0 Mean & F0 Frequency & \multirow{2}{*}{ Literature Comparison (Marx et al.)} \\
& & (ms) & (RMS) & (kHz) & Modulation & \\
\midrule
\multirow{2}{*}{ Development}
& Cluster 0 & 142 ± 47 & 0.025 ± 0.018 & 3.02 ± 0.48 & Ascending modulation (F0 1st diff +37.0 Hz); Rise from onset to F0 peak (F0 slope +7.0 kHz/s); moderate frequency range (F0 bandwidth 855 Hz). & Partial alignment with pleasure calls. Duration: 142ms vs Marx 23--90ms. Energy: 0.025 vs Marx 0.001--0.02. F0: 3.0kHz vs Marx 2.5--5.5kHz. Ascending matches Marx's pleasure calls identified pattern.\\
\cline{2-7}
& Cluster 1 & 346 ± 89 & 0.138 ± 0.057 & 3.33 ± 0.28 & Stable/Flat modulation (F0 1st diff +8.6 Hz); Rapid rise from onset to F0 peak (F0 slope +11.8 kHz/s); High bandwidth suggests complexity (F0 bandwidth 1207 Hz). & Partial alignment with contact calls. Duration: 346ms vs Marx 100--250ms. Energy: 0.138 vs Marx 0.02--2. F0: 3.3kHz vs Marx 2--6.5kHz. Extended duration, stable F0 = contact function.\\
\midrule
\multirow{2}{*}{Control Group} 
& Cluster 0 & 115 ± 33 & 0.097 ± 0.090 & 3.27 ± 0.43 & Slightly Descending modulation (F0 1st diff -2.9 Hz); Minimal upward trend (F0 slope +3.4 kHz/s) ; Low bandwidth modulation (F0 bandwidth 502 Hz) & Partial alignment with short peeps. Duration: 115ms vs Marx 40--100ms. Energy: 0.097 vs Marx 0.004--0.01. F0: 3.3kHz vs Marx 3--5kHz. Descending modulation matches short peep pattern.\\
\cline{2-7}
& Cluster 1 & 186 ± 58 & 0.383 ± 0.271 & 3.78 ± 0.31 & Moderately Ascending frequency (F0 1st diff +39.9 Hz; F0 slope +11.5 kHz/s); High bandwidth = complex (F0 bandwidth 1073 Hz).& Match contact call. Duration: 186ms vs Marx 100--250ms. Energy: 0.383 vs Marx 0.02--2. F0: 3.8kHz vs Marx 2--6.5kHz. Ascending frequency modulation consistent with contact function.\\
\midrule
\multirow{2}{*}{ VPA Group}
& Cluster 0 & 117 ± 32 & 0.070 ± 0.052 & 3.47 ± 0.50 & Near-flat, slightly ascending modulation (F0 1st diff +3.9 Hz); Flattened frequency contour (F0 slope +3.2 kHz/s; F0 bandwidth 488 Hz) & Partial matching short peep. Duration: 117ms vs Marx 40--100ms. Energy: 0.070 vs Marx 0.004--0.01. F0: 3.5kHz vs Marx 3--5kHz. Atypical ascending vs Marx descending pattern.\\
\cline{2-7}
& Cluster 1 & 145 ± 45 & 0.420 ± 0.158 & 3.58 ± 0.35 & Weakly Ascending modulation (F0 1st diff +14.7 Hz; F0 slope +4.7 kHz/s); Narrow frequency range = impaired complexity (F0 bandwidth 551 Hz).& Partial matching contact call. Duration: 145ms vs Marx 100--250ms. Energy: 0.420 vs Marx 0.02--2. F0: 3.6kHz vs Marx 2--6.5kHz. Atypical ascending vs Marx descending; reduced complexity.\\
\bottomrule
\end{tabular}
\caption{Comparison of identified clusters in the development and experimental datasets with call types described from Marx et al. \cite{marx2001vocalisation}}
\label{tab:literature_computational_comparison}
\end{table}
\end{landscape}

\end{document}